\documentclass[fleqn,10pt]{wlscirep}

\usepackage[numbered]{bookmark}
\usepackage{lipsum}
\usepackage[utf8]{inputenc}
\usepackage[T1]{fontenc}
\usepackage[numbers]{natbib}
\usepackage{makecell, caption, booktabs}
\usepackage{textcomp}
\usepackage{siunitx}
\usepackage{subcaption}
\usepackage{listings}
\DeclareCaptionSubType*[Alph]{table}
\DeclareCaptionLabelFormat{mystyle}{Table~\bothIfFirst{#1}{ ̃}#2}

\usepackage{placeins}
\usepackage{float}
\usepackage{csquotes}
\usepackage[tableposition=top]{caption}
\usepackage{adjustbox}
\titlespacing{\equation}{0pt}{*0}{*0}
\titlespacing{\table}{0pt}{*0}{*0}
\titlespacing{\figure}{0pt}{*0}{*0}
\usepackage{xcolor}
\usepackage{tabularx}
\colorlet{tablegray}{gray!50!white}

\newcommand\numberthis{\addtocounter{equation}{1}\tag{\theequation}}
\usepackage{textcomp}
\usepackage{mathpazo}
\usepackage{eulervm}

\DeclareUnicodeCharacter{2212}{\textminus}

\usepackage{siunitx}
\title{An Aggregate Method for Thorax Diseases Classiﬁcation}

\author[1*]{Bayu Adhi Nugroho}
\affil[1]{Curtin University, Bentley, WA, 6102, Australia}
\affil[1]{BayuAdhi.Nugroho@postgrad.curtin.edu.au}

\begin{abstract}
A common problem found in real-word medical image classiﬁcation
is the inherent imbalance of the positive and negative patterns in the dataset where positive patterns are usually rare.  Moreover, in the classiﬁcation of multiple classes with neural network, a training pattern is treated as a positive pattern in one output node and negative in all the remaining output nodes. In this paper, the weights of a training pattern in the loss function are designed based not only on the number of the training patterns in the class but also on the diﬀerent nodes where one of them treats this training pattern as positive and the others treat it as negative. We propose a combined approach of weights calculation algorithm for deep network training and the training optimization from  the state-of-the-art deep network architecture for thorax diseases classiﬁcation problem.  Experimental results on the Chest X-Ray image dataset demonstrate that this new weighting scheme improves classiﬁcation performances, also the training optimization from the EfficientNet improves the performance furthermore. We compare the aggregate method with several performances from the previous study of thorax diseases classiﬁcations to provide the fair comparisons against the proposed method.
\end{abstract}
\begin{document}

\flushbottom
\maketitle
%
%
\thispagestyle{empty}

\pdfbookmark[section]{Introduction}{Introduction}
\section*{Introduction}
\label{sec:introduction}
The traditional classifiers such as an SVM, a decision tree or a logistic regression require feature engineering to perform classification. The better features chosen during the feature engineering will produce more accurate classification performance. Despite the necessity of the feature engineering, a neural network has the advantages of performing end-to-end training to output a final classification's prediction. The removal of feature engineering in the neural network will reduce the risk of the use of incorrect features for classification. To develop a neural network for medical diagnosis, patient data are necessary; however, the positive class is minority and the negative class is majority. The neural network is biased to the majority class and has poor performance on the minority class. The common methods to balance the number of positive and negative class for the traditional classifiers is by the use of undersampling and oversampling approaches. After applying those methods, the numbers of each training-pattern are equal. The more sophisticated approach might involve algorithmic technique to perform cost-sensitive training \cite{Johnson2019}. \enquote{The effective number of samples is defined as the volume of sample}  \cite{DBLP:conf/cvpr/CuiJLSB19}. Cui et al. \cite{DBLP:conf/cvpr/CuiJLSB19} developed a metric to determine the effective samples and reformulated the loss function based on the numbers of effective samples in the positive and negative classes. The work of Wang et al. \cite{wang2017chestxray} contributes to provide the availability of a novel Chest X-Ray dataset. Both \cite{DBLP:conf/ciarp/GundelGGLMC18, wang2017chestxray} use the same formulation to balance the dataset, however the method from\cite{DBLP:conf/ciarp/GundelGGLMC18, wang2017chestxray} is different than \cite{DBLP:conf/cvpr/CuiJLSB19} to deal with the imbalance problem. Cui et al.'s approach \cite{DBLP:conf/cvpr/CuiJLSB19} treats the contributions of training patterns to the loss function equally for all
the output nodes, this is contrary to the Wang and Gundel et al's methods \cite{DBLP:conf/ciarp/GundelGGLMC18, wang2017chestxray} which use the distinct weights from positive and negative classes as the multipliers in the loss-function. Although better classification performance can potentially be 
achieved by Cui et al \cite{DBLP:conf/cvpr/CuiJLSB19}, the approach only addresses effective samples \cite{DBLP:conf/cvpr/CuiJLSB19} and the imbalances of positive-negative classes have not been tackled. In this paper, a novel weights function for focal-loss is proposed to address the imbalance of positive-negative classes,which tackles the classification correctness in both positive and negative samples when training the neural networks. The performance of the proposed focal-loss function is evaluated by performing Chest X-Ray classification which is involved with imbalanced data \cite{wang2017chestxray}. We also propose the use of EfficientNet \cite{pmlr-v97-tan19a} with progressive image resizing under two-stage training in complement with the proposed loss-function. The motivation to use EfficientNet is to inspect the outcome of the proposed loss function into different architecture scaling. The aggregate of the proposed loss-function and the two-stage EficientNet training achieved $2.10\%$ improvement, which is measured with the area under the receiver-operating-characteristic curve (AUROC). Also, heatmap visualization shows that the proposed aggregate approach can achieve better coverage of the diseases. According to Baltruschat et al. \cite{Baltruschat2019} the current state-of-the-art performance for the dataset \cite{wang2017chestxray} classification performance was achieved by G{\"{u}}ndel et al.  \cite{DBLP:conf/ciarp/GundelGGLMC18}. Further research by Guan et al. \cite{GUAN202038} which use three-stage training procedures reported better performance than G{\"{u}}ndel et al. \cite{DBLP:conf/ciarp/GundelGGLMC18}. However, the work \cite{GUAN202038} did not share the split-sets which is critical for the performance evaluation, also the re-implementation by another party in github \cite{ien001} reported lower results. We also notice that the re-implementation \cite{ien001} of \cite{GUAN202038} did not share identical sets with the work of  G{\"{u}}ndel et al.  \cite{DBLP:conf/ciarp/GundelGGLMC18}. Baltruschat et al. \cite{Baltruschat2019} noticed that different split-sets would lead to different performances for the dataset \cite{wang2017chestxray}. To have the fair benchmarks, we report several results from various split-sets options for the performance evaluation. We perform three split-sets experiments configuration setup, which aims to provide better evaluation and  have the comprehensive analysis; the first is by the use of \enquote{official} split from \cite{r_2017}, the second is under five-folds cross-validation configuration which has also been used in the work of Baltruschat et al. \cite{Baltruschat2019}, and the last one is by the use of identical splits from the public Github-page \cite{ien001, arnoweng_2017}. We achieve state-of-the-art results for the classification problem of the Chest X-Ray dataset \cite{wang2017chestxray}, measured under these three split-sets configurations. This research is mainly contributing to the improvement of medical image classification problem. We tackle the imbalance problem within Chest X-Ray dataset and propose the advancement of the use of state-of-the-art neural net architecture for the final classification performance;The EfficientNet with two-stage training. In summary, our contribution is to propose an approach which can combine weights calculation algorithm for deep-network and the optimization of training strategy from the state-of-the-art architecture. The Introduction section of this paper provides a brief introduction and overview of the research. The Method Section mainly discusses the existing classification approach and the proposed method. The Experiments and Results Section presents the results from experiments. The Discussion Section give a more in-depth discussion about the outcome, then The Conclusion Section explains the conclusion from the research. 
\pdfbookmark[section]{Method}{Method}
\section*{Method}
\pdfbookmark[subsection]{The Existing Weights Function and Network Architecture}{The Existing Weights Function and Network Architecture}
\subsection*{The Existing Weights Function and Network Architecture}
Wang et al. \cite{wang2017chestxray} and Gündel et al.  \cite{DBLP:conf/ciarp/GundelGGLMC18} defined the weights, $\omega_{k+}$ and $\omega_{k-}$, of the positive and negative samples for the $k-th$ pattern. 
\begin{align*} 
\numberthis \label{eqn:GundelandWang}
\omega_{k+} = \frac{P_k + N_k}{P_k} 
\\
\omega_{k-} = \frac{P_k + N_k}{N_k}
\end{align*} 
where $P_k$ and $N_k$ are the numbers of positive and negative samples for the the $k^{th}$ pattern. However, Cui et al. \cite{DBLP:conf/cvpr/CuiJLSB19} used both $\omega_{k+}$ and $\omega_{k-}$ equally to develop the loss function. In the manuscript, \cite{wang2017chestxray} and \cite{DBLP:conf/ciarp/GundelGGLMC18} did not use an identical dataset. \cite{wang2017chestxray} used Chest X-Ray 8 which consists of eight classes with 108,948 images. Whilst  \cite{DBLP:conf/ciarp/GundelGGLMC18} used Chest X-Ray 14 which consists of fourteen classes with 121,120 images. However, we can find the results from the method of Wang et al. \cite{wang2017chestxray} under the Chest X-Ray 14 \enquote{official split} configuration in the manuscript of Gündel et al.  \cite{DBLP:conf/ciarp/GundelGGLMC18}. Both works implemented Equation~\ref{eqn:GundelandWang} in the loss-function according to the literature \cite{wang2017chestxray, DBLP:conf/ciarp/GundelGGLMC18}. Therefore, we can conclude the Equation~\ref{eqn:GundelandWang} applies to the samples of the training set. Lin et al. \cite{inproceedings} proposed the focal-loss function:
 \begin{equation}
 \label{eqn:LinFoc}
     L_{foc}(\mathit{p}) = - \alpha ( 1 - \mathit{p}) ^{\gamma } log(\mathit{p}).
 \end{equation}
 $p$ is the prediction. In Equation~\ref{eqn:LinFoc}, the parameter $\alpha$  attempts to balance the positive-negative samples while $\gamma$ adjusted to release the easy samples and to dominate the hard samples, where the easy and hard samples are those classified correctly and incorrectly respectively. Generally,  $\gamma \geq 0$; when $\gamma =  0$ focal-loss is the same as an ordinary cross-entropy loss \cite{inproceedings}. The experimental results showed that the easy samples are down-weighed when $\gamma\approx 1$; The samples are further down-weighed when $\gamma > 1$. 
Determination of $\alpha$ is discussed to demonstrate the impact to the focal loss function (Equation~\ref{eqn:LinFoc}). The parameters chosen as below \cite{DBLP:conf/cvpr/CuiJLSB19}:
\begin{align*} 
\numberthis \label{eqn:Cui1}
\beta = \frac{(N - 1)}{N}
\\
\alpha_{k}(\beta) = \frac{1 - \beta }{1 - \beta ^{n_{k}}}
\\
N(\beta) = \sum \alpha_{k}(\beta)
 \end{align*} 
where $n_{k}$ is the number of the $k_{th}$ pattern, and $N$ is the number of samples. Conceptually, $\beta$ is used to adjust the significance of the number of samples. $N(\beta)$ is the sum of all $\alpha_k$-s which is corresponded to the $\beta$ value for each $k$-pattern. $N(\beta)$ is used for normalization with the number of patterns. However, the work from Cui et al, \cite{DBLP:conf/cvpr/CuiJLSB19} ignores the negative patterns into the weight calculations, this dismissed the very important variables because the negative patterns from negative classes are commonly to find in the medical images classification problem.\\
\textbf{The DenseNet-121 Network}\\
DenseNet-121 is popular to perform classification \cite{wang2017chestxray} with some other methods \cite{Baltruschat2019, DBLP:conf/ciarp/GundelGGLMC18, wang2017chestxray, DBLP:journals/corr/abs-1803-07703} which use ResNet \cite{DBLP:conf/cvpr/HeZRS16}. DenseNet \cite{huang2017densely} and ResNet \cite{DBLP:conf/cvpr/HeZRS16} utilize different skip-connection approaches to pass features from previous layers to later layers. ResNet \cite{DBLP:conf/cvpr/HeZRS16} performs
a summation of features for the skip-connections while DenseNet  \cite{huang2017densely} performs concatenation from features. After the input layer, DenseNet utilises 7x7 convolution in a stride 2 mode and it uses 3x3 max pooling also in stride 2 mode. Then it concatenate features in the first Dense block. There are four Dense blocks in DenseNet, each Dense block at least consists of six consecutive of a 1x1 convolution layer followed by a 3x3 convolution layer. The numbers of these consecutive 1x1 and 3x3 layers in Dense blocks are depend on the types of DenseNet which are either 121,169,201 or 264 layered DenseNet, but all of these DenseNet configurations have four Dense blocks and the differences are only in the number of consecutive convolution layers within a Dense block. The concatenated features from a Dense block in DenseNet are then downsampled through a transition layer. The transition layer consists of a 1x1 convolutional layer and a 2x2 average pool layer in stride 2 mode. A Dense block in DenseNet is followed by a transition layer consecutively. ChexNet by Rajpurkar et al. \cite{DBLP:journals/corr/abs-1711-05225} initiates the popularity of DenseNet-121 as the backbone network to perform the Chest X-Ray classification. ChexNet \cite{DBLP:journals/corr/abs-1711-05225} consists the sigmoid functions in the last layer. ChexNet changes the output dimension of the final classification layer of DenseNet-121 from 1024 dimension of softmax output into 14 dimension of sigmoid functions. The change from 1024 to 14 nodes reflects the number of classification's labels in the Chest X-Ray dataset \cite{wang2017chestxray}. Table~\ref{tab:layerl} depicts the layer-differences between ChexNet \cite{DBLP:journals/corr/abs-1711-05225} and DenseNet \cite{huang2017densely}.
\begin{table*}[!h]
    \centering
    \caption{The Layer Comparison DenseNet-121 and ChexNet}
    \setlength{\tabcolsep}{0.6\tabcolsep}
    \label{tab:layerl} 
    \resizebox{13cm}{!}{
    \begin{tabular}{ | l | l | l  l | l  l | }
\hline
	Layers & Output Size & DenseNet - 121 &  & ChexNet &  \\ \hline
	 & 112 x 112 & 7x7 CONV stride 2 & & 7x7 CONV stride 2 &    \\ 
	 & 56 x 56 &  Max Pool stride 2 &  & Max Pool stride 2  &  \\ \hline
	Dense Block (1) & 56 x 56 & 1 x 1 CONV & x  6 & 1 x 1 CONV & x  6 \\ 
	 &  & 3 x 3 CONV &  & 3 x 3 CONV &  \\ \hline
	Transition (1) & 56 x 56 & 1 x 1 CONV  &  & 1 x 1 CONV  &  \\ 
	 & 28 x 28 & 2 x 2 Avg Pool stride 2  &  &  2 x 2 Avg Pool stride 2 &  \\ \hline
	Dense Block (2) & 28 x 28 & 1 x 1 CONV & X 12 & 1 x 1 CONV & X 12 \\
	 &  & 3 x 3 CONV &  & 3 x 3 CONV &  \\ \hline
	Transition (2) & 28 x 28 & 1 x 1 CONV &  & 1 x 1 CONV   &  \\ 
	 & 14 x 14 & 2 x 2 Avg Pool stride 2 &  & 2 x 2 Avg Pool stride 2 &  \\ \hline
	Dense Block (2) & 14 x 14 & 1 x 1 CONV & x 24 & 1 x 1 CONV & x 24 \\ 
	 &  & 3 x 3 CONV &  & 3 x 3 CONV &  \\ \hline
	Transition (3) & 14 x 14 &  1 x 1 CONV  &  &  1 x 1 CONV  &  \\ 
	 & 7 x 7 & 2 x 2 Avg Pool stride 2 &  & 2 x 2 Avg Pool stride 2  &  \\ \hline
	Dense Block (4) & 7 x 7 & 1 x 1 CONV & x 16 & 1 x 1 CONV & x 16 \\
	 &  & 3 x 3 CONV &  & 3 x 3 CONV &  \\ \hline
	Classification Layer & 1 x 1 & 7 x7 GLOBAL AVG POOL &  & 7 x7 GLOBAL  AVG POOL &  \\ 
	 &  & 1000D   SOFTMAX &  & 14D SIGMOID &  \\ \hline
    \end{tabular}
}
\end{table*}
\pdfbookmark[subsection]{The Proposed  Weights Function and Network Architecture}{The Proposed  Weights Function and Network Architecture}
\subsection*{The Proposed  Weights Function and Network Architecture}
 The normalization of  $\alpha_{k}$ formulated in Equation~\ref{eqn:Norm} is used to weight the $k^{th}$ pattern:
\begin{align*} 
\numberthis \label{eqn:Norm}
     \widetilde{\alpha_{k}}(\beta) = \frac{C}{N(\beta)}\cdot \alpha_{k}(\beta)
 \end{align*} 
where $C$ is the number of class. Although Cui et al \cite{DBLP:conf/cvpr/CuiJLSB19} proposed the grid search to determine $\beta$ based on their formulation, the separable weights of a positive and negative pattern have not been addressed. In this paper, we integrate the separability of positive and negative patterns into the loss-function in order to improve the classification capability of Cui et al's approach. The hypotheses address the importance of both positive and negative pattern weights to improve end-to-end training. 
\begin{align*} 
 \numberthis \label{eqn:Wplus}
\omega_{k+} = \widetilde{\alpha_{k}}(\beta)
 \end{align*}
where $\omega_{k+}$ are the weights for positive samples of the $k^{th}$ pattern. The Equation~\ref{eqn:Wplus} is an elaboration point between \cite{DBLP:conf/cvpr/CuiJLSB19} and our proposed method. We deliberately assign $\alpha_k$ to each sample in $k^{th}$ pattern based on the specified $\omega_{k+}$ weights. The work \cite{DBLP:conf/cvpr/CuiJLSB19} emphasized the importance of effective samples to define the weights and we have two types of weights $\omega_{k+}$ and  $\omega_{k-}$ come  into the proposal. In our proposed approach, $\widetilde{\alpha_{k}}(\beta)$ from \cite{DBLP:conf/cvpr/CuiJLSB19} attempts to determine the weights of only the positively labeled samples, which is given in Equation~\ref{eqn:Wplus}. Also, we determine the weight of the negative patterns:
\begin{align*} 
\numberthis \label{eqn:Wminus}
\omega_{k-} = 1 - \omega_{k+}
 \end{align*} 
Experimental results evaluate the performance of the proposed weights in Equation~\ref{eqn:Wplus} and Equation~\ref{eqn:Wminus} to balance the imbalanced samples.\\
\textbf{The Weighted  Cross Entropy Loss} \\
The formulation for Cross Entropy loss \cite{DBLP:conf/nips/ZhangS18} with the proposed weight is:  
\begin{align*} 
     L_{bce}(\mathit{p})  = \sum_{k=1}^{C} \omega_k \, (-y_{true}^{k}\,log(\mathit{p})) \\
    \omega_k  = \left\{\begin{matrix}
\omega_{k-} & \mathrm{if} & y_{true}^{k} = 0\\ 
\omega_{k+}  & \mathrm{if} & y_{true}^{k} = 1
\end{matrix}%
\right.
\numberthis \label{eqn:bceW}
\end{align*} 
where  $y_{true}^{k}$ are the ground-truth labels for each samples in pattern $k$. To perform the experiments in Section 4, we set the $\omega_{k-} = \omega_{k+} $ for a particular case, the case where we want to see the outcome from Cui et al.'s \cite{DBLP:conf/cvpr/CuiJLSB19} formulation into the dataset \cite{wang2017chestxray} classification problem. The Cross Entropy loss use softmax output by default, whereas the Binary Cross Entropy loss use sigmoid output.\\
\textbf{The Weighted Focal Loss}\\
The formulation for focal loss  with the proposed weight is:  
 \begin{align*} 
     L_{foc}(\mathit{p})  =\sum_{k=1}^{C} \omega_k \, ( - \alpha \, ( 1 - \mathit{p}) ^{\gamma } \, y_{true}^{k} \,  log(\mathit{p})) \\
    \omega_k  = \left\{\begin{matrix}
\omega_{k-} & \mathrm{if} & y_{true}^{k} = 0\\ 
\omega_{k+}  & \mathrm{if} & y_{true}^{k} = 1
\end{matrix}%
\right.
\numberthis \label{eqn:focW}
\end{align*} 
The proposed focal-loss attempts to weight both the easy-hard samples and the positive-negative patterns, which are not addressed by Cui et al.'s approach \cite{DBLP:conf/cvpr/CuiJLSB19}. The proposed focal loss also suits the multiclass classification problem. There is no existing focal-loss method which addresses both effective number of samples and positive-negative patterns weighting.\\
\textbf{The Progressive Image Resizing}\\
Progressive image resizing is the procedure to train a single deep network architecture with incremental input sizes in multiple stages of training. The first stage trains the network with the default image size for the network and then followed by the next stage which utilises the bigger size images and the use of the best performance of the pre-trained model from the previous stage. There is no formal definition of the exact number of steps, but the classification performance will improve to some extents and then saturates, and gain diminishes; this is very specific to the classification problems. We report that the third stage of training with progressive image resizing did not improve the performance of the existing Chest X-Ray classification problem. Another functionality from the progressive-image-resizing is to provide another form of augmentation. It (re)trains the model with the augmentations of different sized-inputs. Several works \cite{DBLP:conf/nips/CaoWGAM19, DBLP:journals/jbd/ShortenK19, 7797091, DBLP:journals/corr/abs-1805-09707} of literature mention that augmentation is a proven method to reduce overfitting. We need to have our final model is risk-free from overfitting, and the two-stage training is our approach to perform that. In summary, we perform the two-stage training to achieve two aims: 1. to improve the classification accuracy, 2. to prevent overfitting.\\ 
\textbf{EfficientNet}\\
The recent work by Tan et al. \cite{pmlr-v97-tan19a} introduced EfficientNet. It proposed a formulation to perform grid-search among three prominent aspects of the deep network's architecture: depth, width and input resolution. The depth defines the number of layers, the width defines the number of nodes for each layer, and the input resolution defines the size of the input images. The compound scaling from those three components are then composed into different architectures from EfficientNet-B0 into EfficientNet-B7. The networks use the mobile inverted bottleneck layers similar to \cite{DBLP:conf/cvpr/TanCPVSHL19, conf/cvpr/SandlerHZZC18}, the layers then concatenated to squeeze-excitation layer \cite{Hu18}. The ReLu6 function is capped at the magnitude of 6; it was used in MobileNetV2 \cite{conf/cvpr/SandlerHZZC18}. However, EfficientNet replaces the use of ReLu6 with Swish. Equation~\ref{eqn:Relu} shows the difference among the ordinary ReLu function, the ReLu6 \cite{Krizhevsky10convolutionaldeep} and the Swish activation function:
\begin{align*} 
\numberthis \label{eqn:Relu}
ReLu(x) & = max(0,x)
\\
ReLu6(x) & = min(max(0,x),6)
\\
Swish(x) & = x \dot ReLu(x)
\end{align*} 
The layers of EfficientNet-B0 are depicted in Table~\ref{tab:layer2}. The further scaling of EfficientNets B0 into B7 are then defined by the grid-search formula as reported in \cite{pmlr-v97-tan19a}. After the input layer the EfficientNet use 3x3 spatial convolutional layer in stride 2 mode, then it uses MBConv1 the linear bottleneck and inverted residual layer \cite{conf/cvpr/SandlerHZZC18}. After the MBconv1 layer the network has six consecutive MBConv6 layers with various 3x3 and 5x5 kernel as listed in Table~\ref{tab:layer2}. Each MBConv6 has three consecutive layers consist of a 1x1 convolutional layer, a 3x3 or 5x5 depth-wise convolutional layer and another 1x1 convolutional layer. Each MBConv1 has two consecutive layers consist of  a 3x3 depth-wise convolutional layer and another 1x1 convolutional layer. The final layer consists of 1x1 convolutional, the global average pooling and 1280 nodes of a fully connected layer. Following the previous modification of DenseNet-121 into the specific implementation of the Chest X-Ray \cite{wang2017chestxray} classification problem, we also modify the final output layer from 1280 nodes into 14 nodes.\\
\begin{table*}[!h]
\centering
    \caption{The EfficientNet-B0 Layer  \cite{pmlr-v97-tan19a}}
    \setlength{\tabcolsep}{0.6\tabcolsep}
    \label{tab:layer2}  
    \resizebox{9.5cm}{!}{
    \begin{tabular}{@{}lllll@{}}
\toprule
Stage & Operator                  & Resolution & Channels & Layers \\ 
\textsubscript{i}     & $\mathcal{F}$\textsubscript{i}                         & $\mathcal{H}$\textsubscript{i} X $\mathcal{W}$\textsubscript{i}       & $\mathcal{C}$\textsubscript{i}        & $\mathcal{L}$\textsubscript{i}      \\ \midrule
1     & Conv 3x3                  & 224x224    & 32       & 1      \\
2     & MBConv1, k3x3             & 112x112    & 16       & 1      \\
3     & MBConv6, k3x3             & 112x112    & 24       & 2      \\
4     & MBConv6, k5x5             & 56x56      & 40       & 2      \\
5     & MBConv6, k3x3             & 28X28      & 80       & 3      \\
6     & MBConv6, k5x5             & 28X28      & 112      & 3      \\
7     & MBConv6, k5x5             & 14x14      & 192      & 4      \\
8     & MBConv6, k3x3             & 7x7        & 320      & 1      \\
9     & Conv 1x1 \& Pooling \& FC & 7x7        & 1280     & 1      \\ \bottomrule
\end{tabular}
}
\end{table*}
\FloatBarrier
\pdfbookmark[subsection]{The Performance Evaluation}{The Performance Evaluation}
\subsection*{The Performance Evaluation}
Suppose we want to have a better perspective about the algorithm performance; we need to apply different metrics to evaluate the results. We apply the AU-PRC (Area Under Precision-Recall Curve) metric for further evaluation; the metric has a different characteristic than AU-ROC (Area Under Receiver-Operating-Characteristics). In the term of baseline, AU-ROC has a fixed baseline of .50 for random classifier and 1 for the perfect classifier, respectively \cite{Hanley1982, DBLP:conf/nips/CortesM03}. In contrast, AU-PRC baseline is dynamic since it heavily depends on the ratio between positive and negative samples \cite{10.1371/journal.pone.0118432}. AU-PRC is more sensitive to data distribution. AU-PRC will have the baseline .50 for a random classifier under the equal number of positive and negative samples; when the number of negative samples ten times than positive samples this baseline will decrease to a smaller number .09 \cite{10.1371/journal.pone.0118432}. The formulation to calculate the baseline of AU-PRC shown in Equation~\ref{eqn:baselineAUPRC} is from the literature \cite{10.1371/journal.pone.0118432}. 
\begin{align*} 
\numberthis \label{eqn:baselineAUPRC}
baseline AUPRC = \frac{positives}{positives + negatives}
 \end{align*}
Suppose we have two classes with an identical value of AU-PRC .50, the interpretation from this particular result will vary for both classes. The .50 AU-PRC is a good result for the class with low positive samples, but it may not be satisfactory for the class with a remarkable number of positive samples. 
\pdfbookmark[section]{Experiments and Results}{Experiments and Results}
\section*{Experiments and Results}
\label{sec:experiments}
The chest X-ray dataset \cite{wang2017chestxray} is used to evaluate the performance of the proposed method. It contains 112,120 Chest X-Ray images from 30,805 unique patients, and it has multilabel of 14 classes diseases. The image resolution is 1024 x 1024 with the 8-bit channel. We downsampled the resolution as 224 x 224 and converted the channel into  RGB which can be adopted to our backbone network. Chest X-Ray 14 only consists of frontal-view images. It does not have any lateral-view cases. The number of positive samples for each class is much less than the negative samples as depicted in Figure~\ref{fig:distribution1} . In our proposed method, the five hyperparameters $\beta$ are given in Equation~\ref{eqn:Betas}.
 \begin{align*} 
 \numberthis \label{eqn:Betas}
     \beta_1 = 1-2.0 \cdot 10^{-6}
;
     \beta_2 = 1-2.0 \cdot 10^{-5}
;
     \beta_3 = 1-2.0 \cdot 10^{-4}
;
     \beta_4 = 1-7.0 \cdot 10^{-4}
;
     \beta_5 = 1-2.0 \cdot 10^{-3}
 \end{align*} 
where $\beta_2$ is determined by Equation~\ref{eqn:Cui1}. The grid-search determines the other $\beta$-s. In the exception of the $\beta_4$, the grid search was performed by changing the $\beta$ value with standard deviation of 10 from $\beta_2$. The current value of $\beta_4$ was chosen because that magnitude is the median between $\beta_3$  and $\beta_5$. Also the results obtained by the proposed method is compared to those obtained by the other six methods, Wang et al. \cite{wang2017chestxray}, Yao et al. \cite{DBLP:journals/corr/abs-1803-07703}, baseline ChexNet \cite{DBLP:journals/corr/abs-1711-05225}, weighted binary cross entropy loss, Balturschat et al. \cite{Baltruschat2019} and G{\"{u}}ndel et al. \cite{DBLP:conf/ciarp/GundelGGLMC18}. The comparison is depicted in Table~\ref{tab:results}.
\pdfbookmark[subsection]{Backbone Network Training}{Backbone Network Training}
\subsection*{Backbone Network Training}
\label{backbone}
\begin{figure}[!h]
	\caption{Training Distribution from Official Split}
	\centering
 	\includegraphics[width=0.65\linewidth]{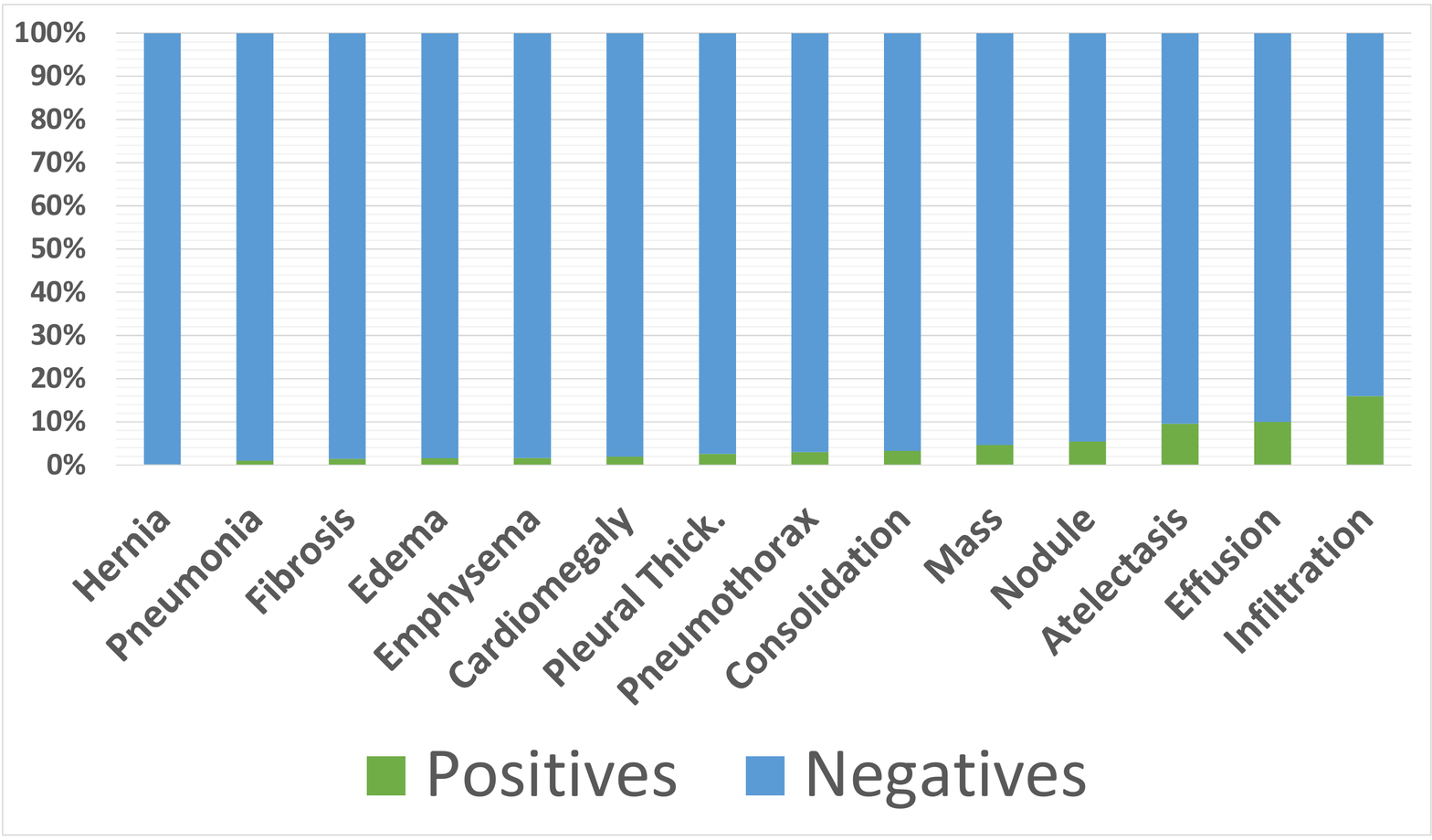}
	\label{fig:distribution1}
\end{figure}

 Since we use the DenseNet 121 \cite{huang2017densely} as the primary backbone network, the availability of pre-trained ImageNet can be used for the classification. Here we used the pre-trained weights from Imagenet to develop the network. The dataset \cite{wang2017chestxray} for base metrics, including the same training, validation, test splitting set from \cite{r_2017}.We refer the split-set \cite{r_2017} as \enquote{official split}. \cite{r_2017} has two groundtruth files as label, they are train\_val\_list.txt which consists of 86524 samples and test\_list.txt  which consists of 25596 samples. Baltruschat et al. \cite{Baltruschat2019} emphasized that different splitting of dataset \cite{wang2017chestxray}  has significantly impact to the classification performance. Since the splitting of training and test data is exactly the same, the benchmark is fair. Figure~\ref{fig:distribution1} and Table~\ref{tab:imbSamples} show that the class distribution is imbalance since the positive and negative samples are very different. We use a single Titan V with 12 Gb GPU memory to develop the network, where 24 hours with 25 epochs of training are required. We only train the Densenet-121 in single-stage training cycle and do not perform progressive image resizing.
\begin{table}[!h]
\centering
\caption{The Imbalance Number of Samples}
\label{tab:imbSamples}
\resizebox{4cm}{!}{
\begin{tabular}{|l|l|}
\toprule
Samples & Number \\ 
\midrule

Healthy        & 60361 \\

\hline
Hernia         & 227   \\
\hline

Pneumonia      & 1431  \\
\hline

Fibrosis       & 1686  \\
\hline

Edema          & 2303  \\
\hline

Emphysema      & 2516  \\
\hline

Cardiomegaly   & 2776  \\
\hline

Pleural Thick. & 3385  \\
\hline

Consolidation  & 4667  \\
\hline

Pneumothorax   & 5302  \\
\hline

Mass           & 5782  \\
\hline

Nodule         & 6331  \\
\hline

Atelectasis    & 11559 \\
\hline

Effusion       & 13317 \\
\hline

Infiltration   & 19894 \\
\bottomrule
\end{tabular}
}
\end{table}
Because we aim to improve the overall classification performance, we also modify the architecture of backbone network from DenseNet-121 into EfficientNet \cite{pmlr-v97-tan19a}. The approach is mainly to expand the performances from the proposed cost-sensitive loss function into better architecture. We are limited only to the use of the EfficientNet-B0 and the EfficientNet-B3 networks for the experiments. This is mainly because we have achieved the peak of the computing-resources limits, and to perform the experiments over all the EfficientNet architectures are not feasible at this stage. Consecutive EfficientNets training requires extensive computations; due the scaling of the image sizes, the depth and also the width of the network. In the other hand, the approach of progressive image resizing only take into account the aspect of image sizes into computational resources; it ignores the depth and the width of the network. To train the EfficientNets, we use the Tesla v100 with 32 Gb of GPU memory. For each network, we performed the two-stage training procedure with progressive-image-resizing as previously discussed. On the first stage we train the network with pre-trained model from ImageNet, then on the second stage we train the network with the best performing model from first stage. The important finetune is the size of the image input; the first stage we use the default input size from the network then we doubled the input size on the second stage. This has been implemented with size of 224 x 224 on the first stage of  EfficientNet-B0 and 448 x 448 on the second stage EfficientNet-B0, also  300 x 300 on the first stage of  EfficientNet-B3 and 600 x 600 on the second stage EfficientNet-B3. We reduce the batch size into half, from 32 on the first stage to 16 on the second stage. The reduced batch size is mainly to ensure the batched-images for each step on each epoch will fit into the GPU's memory boundary. The two-stage training with progressive-image-resizing has successfully improved \textpm $1\%$ to the classification outputs between first stage and second stage for each model.
\pdfbookmark[subsection]{The Baseline Network}{The Baseline Network}
\subsection*{The Baseline Network}
\label{baseline}
We reproduce ChexNet \cite{DBLP:journals/corr/abs-1711-05225} based on \cite{chou_2018}. The experiments performed by our proposed method and the other methods \cite{Baltruschat2019, DBLP:conf/ciarp/GundelGGLMC18, wang2017chestxray} are based on the training and test split in \cite{r_2017} are reported in Table~\ref{tab:results}. However, Rajpurkar et al. \cite{DBLP:journals/corr/abs-1711-05225} never share the split-set for the public. The use of official split  \cite{r_2017} results in lower performance than reported in Rajpurkar et al. \cite{DBLP:journals/corr/abs-1711-05225}. We use the ADAM optimizer as in \cite{DBLP:journals/corr/abs-1711-05225} to develop the neural network of which the optimization is converged at epoch 11. Other researches also used ADAM  \cite{Baltruschat2019, DBLP:conf/ciarp/GundelGGLMC18} and stochastic gradient descent \cite{wang2017chestxray}.
\subsection*{Weighted Binary Cross Entropy with Effective Number of Samples}
This experiment is an adoption of Cui et al.'s \cite{DBLP:conf/cvpr/CuiJLSB19} method into the Chest X-Ray dataset \cite{wang2017chestxray} classification problem. In Cui et al.'s approach \cite{DBLP:conf/cvpr/CuiJLSB19}, the balanced weights between positive and negative is not used; the weights are computed based on the effectiveness number of samples. Cui et al. \cite{DBLP:conf/cvpr/CuiJLSB19}'s used the Equation~\ref{eqn:Cui1} to compute the weights. We perform this experiment to provide evidence of performances which come from \cite{DBLP:conf/cvpr/CuiJLSB19} versus the one come from our approach.  In this experiment, we use binary-cross-entropy as loss function and combine with the weighting into the loss-function. We set the $\omega_{k-} = \omega_{k+} $ for the implementation of Equation~\ref{eqn:bceW} for this particular case, since \cite{DBLP:conf/cvpr/CuiJLSB19} ignores the balanced positives-negatives. The best performance classification for the model is \textbf{also} achieved on epoch 11, similar to \ref{baseline} Baseline. The comparison results with our other experiments are shown in Table~\ref{tab:results}. This method perform only slightly better than the baseline with the $79.24\%$ area under ROC curve.
\pdfbookmark[subsection]{Weighted Focal Loss with Positive and Negative Pattern}{Weighted Focal Loss with Positive and Negative Pattern}
\subsection*{Weighted Focal Loss with Positive and Negative Pattern}
\label{ssec:wFocLoss}
In this experiment, we use the loss function \cite{inproceedings} which is integrated with the focal loss and the proposed weighting. We choose the value of  $\alpha$ value based on \cite{inproceedings} which is between $\left[.25,.75\right]$; we found that $\alpha =  0.5 $ and $\gamma = 1 $ is the best of focal-loss hyperparameters for our proposed method. We use the RANGER (Rectified Adam and LookAhead) optimizer, which requires a smaller number of training epochs to converge. The optimizer converges at epoch 5. We deliberately assign the two-stage training to prevent overfitting and also to improve the performance. This method achieves $82.32\%$ area under ROC curve with two-stage DenseNet-121 and  $83.13\%$ with two-stage EfficientNet-B3. The training time took 71 minutes for one epoch on the first stage with 32 batch size, 180 minutes for one epoch on the second stage with 16 batch size. The test-time took 15 minutes with 8 batch size for the first-stage and took 27 minutes with 8 batch size for the second-stage.

\begin{table*}[!h]
\centering
\caption{Comparison Results with Previous Study under The Official Split}  \label{tab:results}
\centering
\small\addtolength{\tabcolsep}{-4pt}
\begin{tabular}{@{}llllllllllcc@{}}
\toprule
& \makecell{Wang et al.\\ \cite{wang2017chestxray}} & \makecell{Yao \\  et al.\\ \cite{DBLP:journals/corr/abs-1803-07703}  } &  \makecell{baseline \\ reproduce \\ ChexNet \\ \cite{DBLP:journals/corr/abs-1711-05225}} & \makecell{weighted \\  binary \\ cross entropy \\ loss} & \makecell{Baltruschat \\  et al. \\ \cite{Baltruschat2019}}  & \makecell{ G{\"{u}}ndel\\et al. \\ \cite{DBLP:conf/ciarp/GundelGGLMC18}}  &   \makecell{weighted \\  focal \\ loss \\  $\beta =  0.9998$ \\ DenseNet-121  \\ two-stage } & \makecell{weighted \\  focal \\ loss \\  $\beta = 0.9998$ \\ EfficientNet-B3 \\ two-stage}    \\  \midrule
Atelectasis & 0.700 & 0.733 & 0.7541 & 0.7625 & 0.763 & 0.767  &  0.7820 &  0.7919   \\
Cardiomegaly & 0.810 & 0.856 & 0.8787 & 0.8812 & 0.875 & 0.883     & 0.8845 & 0.8917     \\
Effusion & 0.759 & 0.806 & 0.8236 & 0.8266 & 0.822 & 0.806  &    0.8380 & 0.8414    \\
Infiltration & 0.661 & 0.673 & 0.6928 & 0.6939 & 0.694 & 0.709    & 0.7022 & 0.7051      \\
Mass            & 0.693 & 0.777 & 0.8053 & 0.8023 & 0.820 & 0.821    & 0.8329 & 0.8356     \\
Nodule          & 0.669 & 0.724 &  0.7318 & 0.7383 & 0.747 & 0.758   & 0.7863 & 0.8036  \\
Pneumonia       & 0.658 & 0.684 &   0.6980 & 0.7019 & 0.714 & 0.731   & 0.7338 &  0.7366 \\
Pneumothorax    & 0.799 & 0.805 & 0.8378 & 0.8344 & 0.819 & 0.846     & 0.8706 & 0.8909 \\
Consolidation   &  0.703 & 0.711 & 0.7349 & 0.7390 & 0.749 & 0.745   & 0.7537 & 0.7601 \\
Edema          & 0.805 & 0.806 &  0.8345 & 0.8305 & 0.846 & 0.835      &  0.8534 & 0.8609   \\
Emphysema     &   0.833 & 0.842 & 0.8666 & 0.8701 & 0.895 & 0.895     & 0.9413 & 0.9424   \\
Fibrosis      & 0.786 & 0.743 & 0.7957 & 0.8040  & 0.816 & 0.818  & 0.8229 & 0.8408  \\
Pleural Thickening & 0.684 & 0.724 &  0.7456 & 0.7502 & 0.763 & 0.761    & 0.7970  & 0.8080   \\
Hernia          & 0.872  & 0.775  & 0.8684 & 0.8589 & 0.937 & 0.896     & 0.9260 & 0.9286   \\
       
Average & 0.745 & 0.761 & 0.7906 & 0.7924 & 0.806 & 0.807     & 0.8232 & 0.8313\\ \bottomrule
\end{tabular}
\end{table*}
\FloatBarrier
\pdfbookmark[subsection]{The Intuitive Theoretical Background }{The Intuitive Theoretical Background}
\subsection*{The Intuitive Theoretical Background and The Evidence from Experiment}
\begin{table*}[!h]
\caption{The Improvement of The Proposed Weight Calculation}
\label{tab:weightedFocalLoss}
    \begin{tabular}{@{}llllll@{}}
\toprule
 Pathology                 &   \multicolumn{2}{c}{ \makecell{focal \\ loss \\ DenseNet-121}} 
  & \multicolumn{2}{c}{ \makecell{weighted \\  focal \\ loss \\  $\beta =$ \\ 0.9998 \\ DenseNet-121}}
 \\ \midrule
                 &  Validation  & Test & Validation & Test & \\
\hline
Cardiomegaly     &   0.9155 & 0.9096 & 0.9092 & 0.9090\\

Emphysema        &  0.9140 & 0.9178 & 0.9056 & 0.9327\\

Edema            &  0.9141 & 0.8851 & 0.9147 & 0.8917 \\

Hernia           & 0.8614 &  0.9135 & 0.9067 & 0.9404 \\

Pneumothorax     &  0.8896 & 0.8663 & 0.8973 & 0.8749 \\

Effusion         &  0.8822 & 0.8762 & 0.8792 &  0.8827 \\

Mass             & 0.8622 & 0.8430 & 0.8655 & 0.8514  \\

Fibrosis         &  0.8277 &  0.8219 & 0.8313 & 0.8308 \\

Atelectasis      & 0.8191 & 0.8079 & 0.8228 & 0.8259 \\

Consolidation    &  0.8247 & 0.8007 & 0.8224 & 0.8043 \\

Pleural Thicken. & 0.8219 & 0.7874 & 0.8214 & 0.7910 \\

Nodule           &  0.7823 & 0.7751 & 0.7888 & 0.7756  \\

Pneumonia        & 0.7722 & 0.7504 & 0.7586 &  0.7698 \\

Infiltration     & 0.7061 & 0.7073 & 0.7113 &  0.7166 \\

Average          &   0.8424 & 0.8330 & 0.8453 & 0.8427 \\ \bottomrule
\end{tabular}
\end{table*}
\FloatBarrier
Since part of our approach inherits the strength of focal loss \cite{inproceedings}, and also the class-balanced approach \cite{DBLP:conf/cvpr/CuiJLSB19}. We can have further theoretical analysis from the proposed approach intuitively based on \cite{inproceedings} and \cite{DBLP:conf/cvpr/CuiJLSB19}. The main distinction of focal-loss with binary cross entropy loss is the existence of $\alpha$ and $\gamma$ parameter. Cui et al. mention on the paper: \enquote{the class-balanced term can be viewed as an explicit way to set $\alpha$ in focal loss based on the effective number of samples} \cite{DBLP:conf/cvpr/CuiJLSB19}. However, Lin et al. also mention \enquote{a common method for addressing class imbalance is to introduce a weighting factor $\alpha \in  \left [ 1,0 \right ]$ for class 1 and ${1 − \alpha}$ for ${class − 1}$}\cite{inproceedings}. We can inference those two statements into our elaboration in Equation~\ref{eqn:Wminus}. The experiments provide a further evidence for the theory. The improvement from  the change of the proposed formula is \textpm $1\%$ under the test-set according to the Table~\ref{tab:weightedFocalLoss}. Both experiments were performed with $\alpha=0.5$ and $\gamma=1.0$ for the Focal Loss's parameters. Both Table~\ref{tab:weightedFocalLoss} and Table~\ref{tab:agCNN} use an identical split \cite{ien001}. The training-set consists of 78468 images, the validation-sets consists of 11219 images, the test-sets consists of 22433 images. The training time took 38 minutes for one epoch with 32 batch size and the test time took 12 minutes with 8 batch size.

\pdfbookmark[subsection]{The Imbalance Metric Evaluation}{The Imbalance Metric Evaluation}
\subsection*{The Imbalance Metric Evaluation}
\begin{figure}[!h]
	\caption{The Area Under Precision-Recall Curve}
	\centering
 	\includegraphics[width=0.90\linewidth]{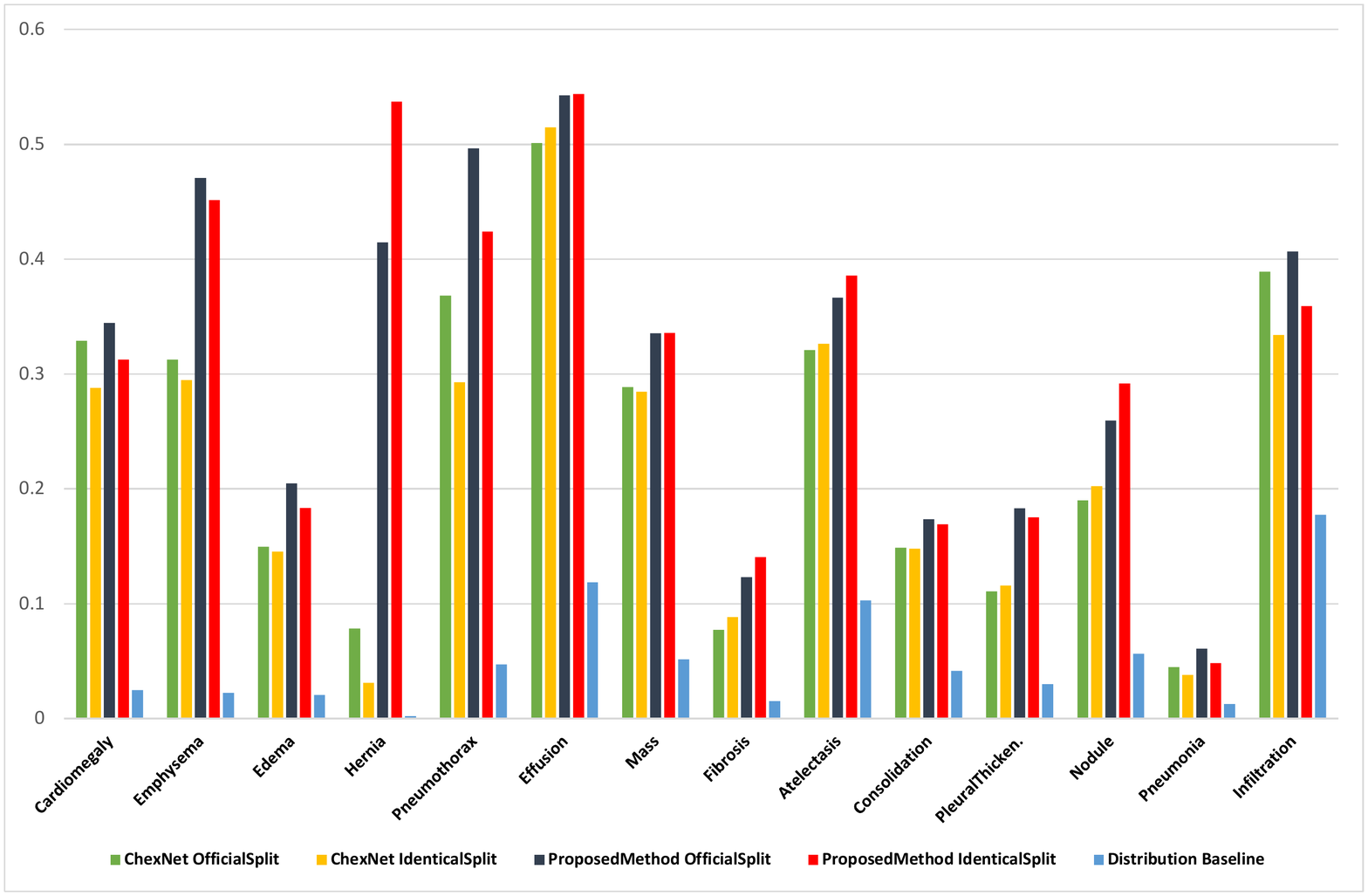}
	\label{fig:auprc}
\end{figure}

\begin{table*}[!h]
\caption{The AU-PRC Improvement}
\label{tab:auPRC}
    \begin{tabular}{@{}lllllll@{}}
\toprule
 Pathology                 &   \multicolumn{2}{c}{ \makecell{reproduce \\ ChexNet \\ \cite{DBLP:journals/corr/abs-1711-05225}}} 
  & \multicolumn{2}{c}{ \makecell{weighted \\  focal \\ loss \\  $\beta =$ \\ 0.9998 \\ EfficientNet-B3 \\ two-stage}} & \multicolumn{2}{c}{ \makecell{baseline \\ AU-PRC \\ from \\ Distribution}}
 \\ \midrule
          Split       &  Official  & \makecell{Identical \\  \cite{ien001}} & Official & \makecell{Identical \\  \cite{ien001}} & \\
\hline
Cardiomegaly     & 0.3288 & 0.2880 & \textbf{0.3444} & \textbf{0.3127} &  0.0247\\

Emphysema          & 0.3125 &  0.2948 & \textbf{0.4706} & \textbf{0.4515} & 0.0224\\

Edema           & 0.1497  &  0.1455 & \textbf{0.2048}  & \textbf{0.1835} & 0.0205\\

Hernia            & 0.0785  & 0.0311 & \textbf{0.4147} & \textbf{0.5372} & 0.0020\\

Pneumothorax      & 0.3683 & 0.2929 & \textbf{0.4965} & \textbf{0.4242} & 0.0472 \\

Effusion          & 0.5012  &  0.5149 & \textbf{0.5428}  & \textbf{0.5439} & 0.1187 \\

Mass         & 0.2887 &  0.2847 & \textbf{0.3355} & \textbf{0.3357} & 0.0515 \\

Fibrosis         & 0.0773 & 0.0886  & \textbf{0.1231} & \textbf{0.1405} & 0.0150\\

Atelectasis        & 0.3208  & 0.3262 & \textbf{0.3664} & \textbf{0.3859} & 0.1030\\

Consolidation     & 0.1488 &  0.1479 & \textbf{0.1736} & \textbf{0.1692} & 0.0416\\

Pleural Thicken.   & 0.1109 &  0.1159  & \textbf{0.1831}  & \textbf{0.1754} & 0.0301\\

Nodule            & 0.1899 &  0.2025  & \textbf{0.2595} &\textbf{0.2919} & 0.0564\\

Pneumonia       & 0.0448  &  0.0381  & \textbf{0.0609} & \textbf{0.0484} & 0.0127\\

Infiltration       &   0.3891 &  0.3342  & \textbf{0.4067} &\textbf{0.3592} & 0.1774\\

Average           & 0.2364 & 0.2218 & \textbf{0.3130} & \textbf{0.3114} & 0.0517 \\\bottomrule
\end{tabular}
\end{table*}
\FloatBarrier
Table~\ref{tab:auPRC} and Figure~\ref{fig:auprc} show the advancement from the proposed method in comparison with previous work \cite{DBLP:journals/corr/abs-1711-05225} also with the baseline retrieved from the dataset. We calculate the baseline of AU-PRC metric directly from the dataset's distribution of positives and negatives samples with the use of Equation~\ref{eqn:baselineAUPRC}. The bold fonts show the top-scores achieved between a same split-set configuration. The hernia has the lowest number of positive samples in the distribution. Despite being the most minority, the proposed algorithm results from hernia a couple of hundred higher AU-PRC than the baseline; shown in Table~\ref{tab:auPRC} and Figure~\ref{fig:auprc}. 
\FloatBarrier
\pdfbookmark[section]{Discussion}{Discussion}
\section*{Discussion}
In order to provide more insights of the effect from different splits into the classification performance, several split-sets has been taken into performance evaluation. The standard procedure is to follow the \enquote{official} splits \cite{r_2017} and we report the results in Table~\ref{tab:results}. To the best of our knowledge, only \cite{Baltruschat2019} reported the performance evaluation of a random  five-folds cross validation from the Chest X-Ray dataset \cite{wang2017chestxray} and we report the results from the proposed method in Table~\ref{tab:fiveFolds}.  There are other split-sets which are considered \enquote{non-standard} settings, these splits are from github pages. \cite{ien001} is the third party re -implementation of \cite{GUAN202038}, and also \cite{arnoweng_2017} is the third party re-implementation of \cite{DBLP:journals/corr/abs-1711-05225}. However, after further investigation \cite{ien001} and \cite{arnoweng_2017}  are under the identical training, validation and testing sets. We report the results with the custom-sets \cite{ien001, arnoweng_2017} in Table~\ref{tab:agCNN}. The results in Table~\ref{tab:rate} are the improvements made in compare with the most recent research \cite{GUAN202038}. The one-by-one comparison for each disease with latest research \cite{GUAN202038} as listed in Table~\ref{tab:agCNN}. We achieve better performances in compare with the work of Guan et. al \cite{GUAN202038} and we propose technically more simple approach to achieve the results. Since the diversity of split-sets is a well-known problem for the dataset's \cite{wang2017chestxray} evaluation, the use cross validation is a fair method to follow. \cite{Baltruschat2019} is the \textbf{only} work that reported performing cross-validation to the dataset \cite{wang2017chestxray}, we achieve better performance in 5-folds cross validation experiment than the work of Baltruschat et. al \cite{Baltruschat2019}. The class-activation-mapping (CAM) method \cite{7780688,DBLP:conf/cvpr/YangH0CHZ19} visualizes the discriminative-features from the deep-network's last layer in the form of heatmap localisation, the more heatmap visualization match the groundtruth bounding-box from dataset means the network has better understanding of the images. We visualize our classification performances with heatmap from CAM method in Table~\ref{tab:Heatmap}. We obtain the bounding-boxes as the annotation groundtruth for only 8 (eight) classes which are available from the file BBox\_List \_2017.csv \cite{r_2017}. The annotations consists of 984 images, and the number of samples for each class is not distributed evenly. Table~\ref{tab:Heatmap} shows that the networks which are equipped with the proposed method read the area of the disease better than the baseline. 

\begin{table*}
    \begin{minipage}{0.45\textwidth}
       \caption{Results from Five-Folds Cross-Validation}  
       \label{tab:fiveFolds}
\begin{tabular}{@{}lll@{}}
\toprule
Pathology        & \makecell{Baltruschat \\  et al. \\ \cite{Baltruschat2019}}         & \makecell{weighted \\  focal \\ loss \\  $ \beta $  = \\ 0.9998 \\ EfficientNet-B3 \\ two-stage}          \\ \midrule
Cardiomegaly     & 89.8 \textpm 0.8 & 90.6 \textpm 2.4 \\
Emphysema        & 89.1 \textpm 1.2 & 94.6 \textpm 1.2 \\
Edema            & 88.9 \textpm 0.3 & 90.3 \textpm 0.9 \\
Hernia           & 89.6 \textpm 4.4 & 92 \textpm 1.3    \\
Penumothorax     & 85.9 \textpm 1.1 & 91.2 \textpm 1.2  \\
Effusion         & 87.3 \textpm 0.3 & 88.5 \textpm 0.5  \\
Mass             & 83.2 \textpm 0.3 & 86.9 \textpm 1.1   \\
Fibrosis         & 78.9 \textpm 0.5 & 82.2 \textpm 2.5  \\
Atelectasis      & 79.1 \textpm 0.4 & 83.3 \textpm 0.7  \\
Consolidation    & 80.0 \textpm 0.7 & 80.9 \textpm0.5   \\
Pleural Thicken. & 77.1 \textpm 1.3 & 82.7 \textpm1.3   \\
Nodule           & 75.8 \textpm 1.4 & 81.7 \textpm 1.4  \\
Pneumonia        & 76.7 \textpm 1.5 & 77 \textpm 1.9    \\
Infiltration     & 70.0 \textpm 0.7 & 72.8 \textpm 4.5   \\
Average          & 82.2 \textpm 1.1 & 85.3 \textpm 0.6  \\ \bottomrule
\end{tabular}
    \end{minipage}
    \hfill
    \begin{minipage}{.55\linewidth}
  \setlength{\tabcolsep}{0.6\tabcolsep}
\caption{Identical Split Comparison \cite{ien001}} \label{tab:agCNN} 
 \resizebox{9cm}{!}{
\begin{tabular}{@{}lllllll@{}}
\toprule
 Pathology                 &  \makecell{ third party \cite{ien001} \\ of \\ Guan  et al. \cite{GUAN202038}}  & \multicolumn{2}{c}{ \makecell{weighted \\  focal \\ loss \\  $\beta =$ \\ 0.9998 \\ DenseNet-121}} 
  & \multicolumn{2}{c}{ \makecell{weighted \\  focal \\ loss \\  $\beta =$ \\ 0.9998 \\ EfficientNet-B3}}
 \\ \midrule
                 &       & stage 1  & stage 2 & stage 1 & stage 2 & \\
\hline
Cardiomegaly     & 0.9097 &  0.9090 &  0.9142 & 0.9137 & 0.9144 \\

Emphysema        & 0.8905 & 0.9327 & 0.9425 & 0.9471 & 0.9558 \\

Edema            & 0.9185 & 0.8917 & 0.8930 & 0.9021 & 0.9071 \\

Hernia           & 0.9064 & 0.9404 & 0.9427 & 0.9357 & 0.9409 \\

Pneumothorax     & 0.8794 & 0.8749 & 0.8854 & 0.9003 & 0.9092 \\

Effusion         & 0.8843 & 0.8827 & 0.8885 & 0.8899 & 0.8923 \\

Mass             & 0.8707 & 0.8514 & 0.8568 & 0.8596 & 0.8669 \\

Fibrosis         & 0.8208 & 0.8308 & 0.8458 & 0.8526 & 0.8657 \\

Atelectasis      & 0.8225 & 0.8259 & 0.8307 & 0.8350 & 0.8397 \\

Consolidation    & 0.8210 & 0.8043 & 0.8115 & 0.8124 & 0.8208 \\

Pleural Thicken. & 0.8127 & 0.7910 & 0.8011 & 0.8041 & 0.8136 \\

Nodule           & 0.7691 &  0.7756 & 0.8144 & 0.8043 & 0.8293 \\

Pneumonia        & 0.7614 & 0.7698 & 0.7726 & 0.7721 & 0.7703 \\

Infiltration     & 0.7006 & 0.7166 & 0.7178 & 0.7297 & 0.7363 \\

Average          & 0.8405 & 0.8426 & 0.8512 & 0.8542 & 0.8616 \\ \bottomrule
\end{tabular}
}

\small
We  found the third party re-implementation \cite{ien001} 
reported lower performances than reported in the paper \cite{GUAN202038}. 

Guan et al. \cite{GUAN202038} did not provide the official code and  split-sets. 
The critical classification problems for the dataset \cite{wang2017chestxray}, different splits will lead to different performances \cite{Baltruschat2019}.
    \end{minipage}
\end{table*}

\begin{table*}[!h]
\caption{The Improvement Rate}
\label{tab:rate}
\centering
\begin{subtable}{\textwidth}
\small
\caption{}
\resizebox{17cm}{!}{
\begin{tabular}{|l|l|l|l|l|l|l|l|l|l|}
\hline
\textbf{Name} & 
\textbf{Hernia} & 
\textbf{Pneumonia} & 
\textbf{Fibrosis}   & 
\textbf{Edema}   & 
\textbf{Emphysema} & 
\textbf{Cardiomegaly}  & 
\textbf{Pleural Thick.} &  
\textbf{Pneumothorax}  \\
\hline
Rate & $+ 3.45 \%$ & $+ 0.89 \%$  & $+ 4.49 \%$ & $- 1.14 \%$ & $+ 6.53 \%$  & $+ 0.47 \%$ & $+ 0.09 \%$ & $+ 2.98 \%$  \\
\hline
\end{tabular}
}
\end{subtable}
\bigskip
\begin{subtable}{\textwidth}
\tiny
\caption{}
\resizebox{17cm}{!}{
\begin{tabular}{|l|l|l|l|l|l|l|l|l|l|}
\hline
Consolidation & Mass  &  Nodule &  Atelectasis & Effusion & Infiltration & \textbf{Average}  \\
\hline
$- 0.02 \%$ &  $- 0.38 \%$ & $+ 6.02 \%$ & $+ 1.72 \%$   &  $+ 0.80 \%$   &  $+ 3.57 \%$ & \textbf{$+ 2.10 \%$} \\
\hline
\end{tabular}
}
\end{subtable}
\end{table*} 
\FloatBarrier
\begin{table*}[!hb]
             \caption{The Heatmap from Different Methods}
        \label{tab:Heatmap}
        \resizebox{11cm}{!}{
         \begin{tabular}{lccc}
           \toprule
              Pathology &  \makecell{baseline \\ reproduce \\ ChexNet \\ \cite{DBLP:journals/corr/abs-1711-05225}}  &\makecell{weighted \\  focal \\ loss \\ DenseNet-121} & \makecell{weighted \\  focal \\ loss \\  EfficientNet-B3} \\
            \midrule
            Atelectasis & \includegraphics[width=0.090\textwidth]{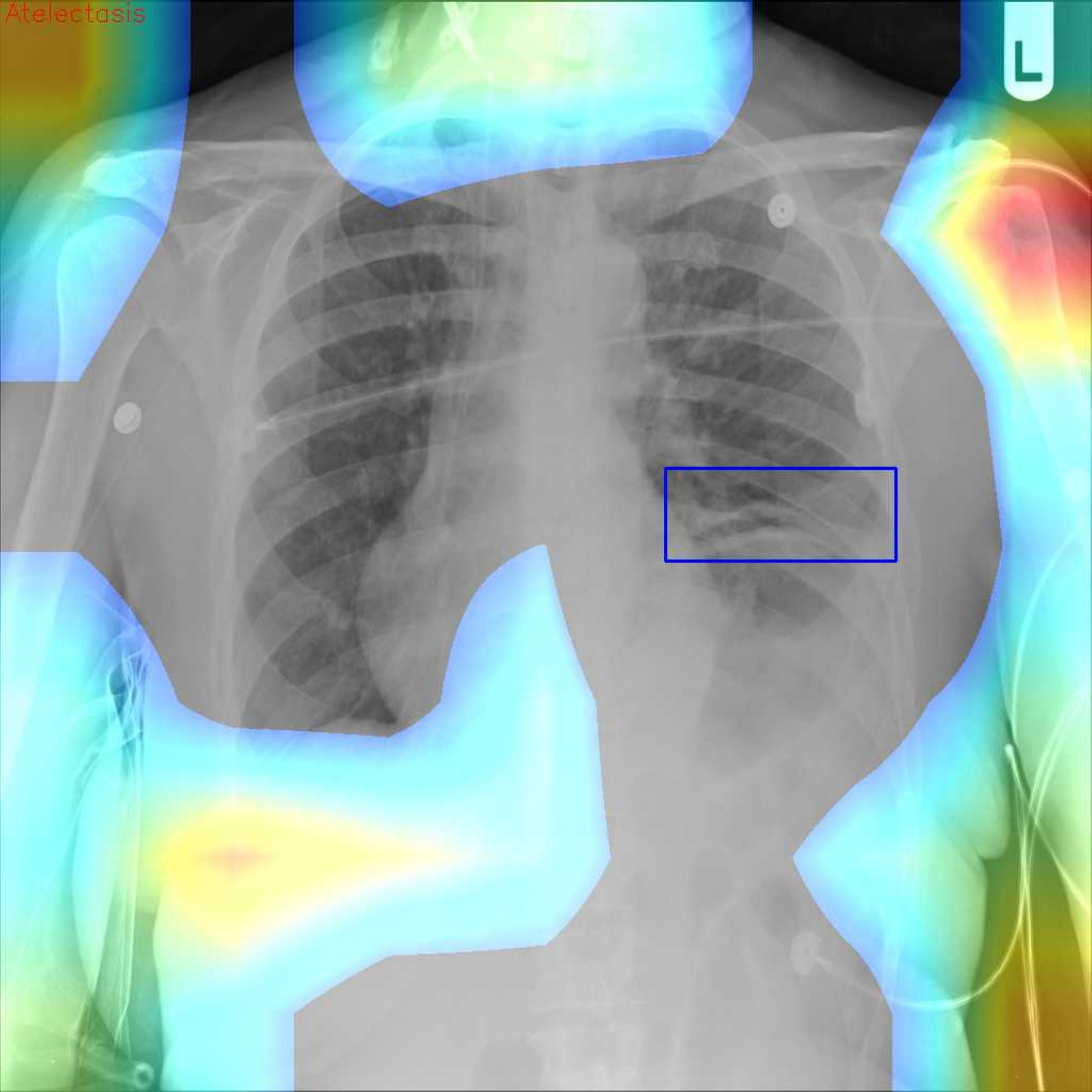} & \includegraphics[width=0.090\textwidth]{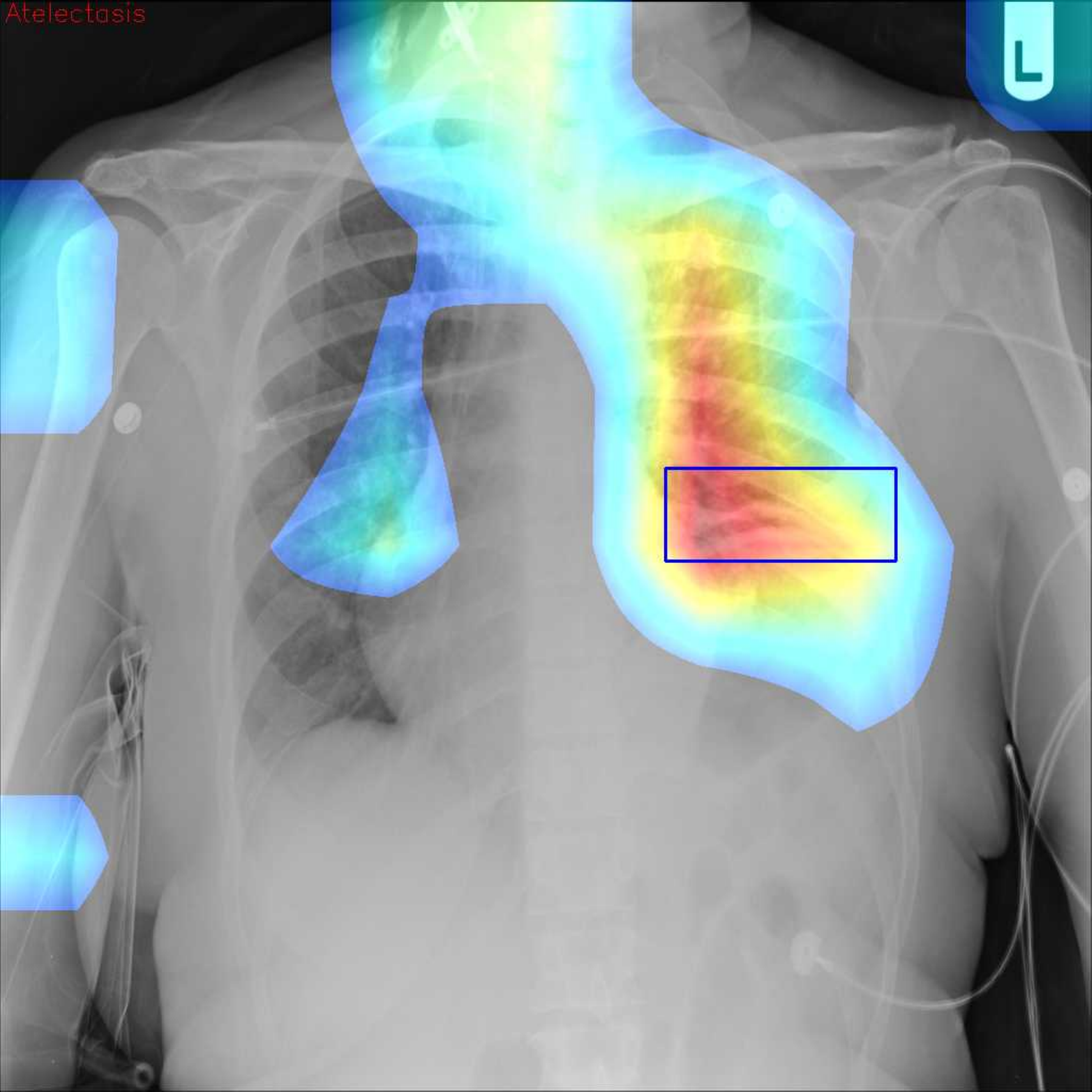} & \includegraphics[width=0.090\textwidth]{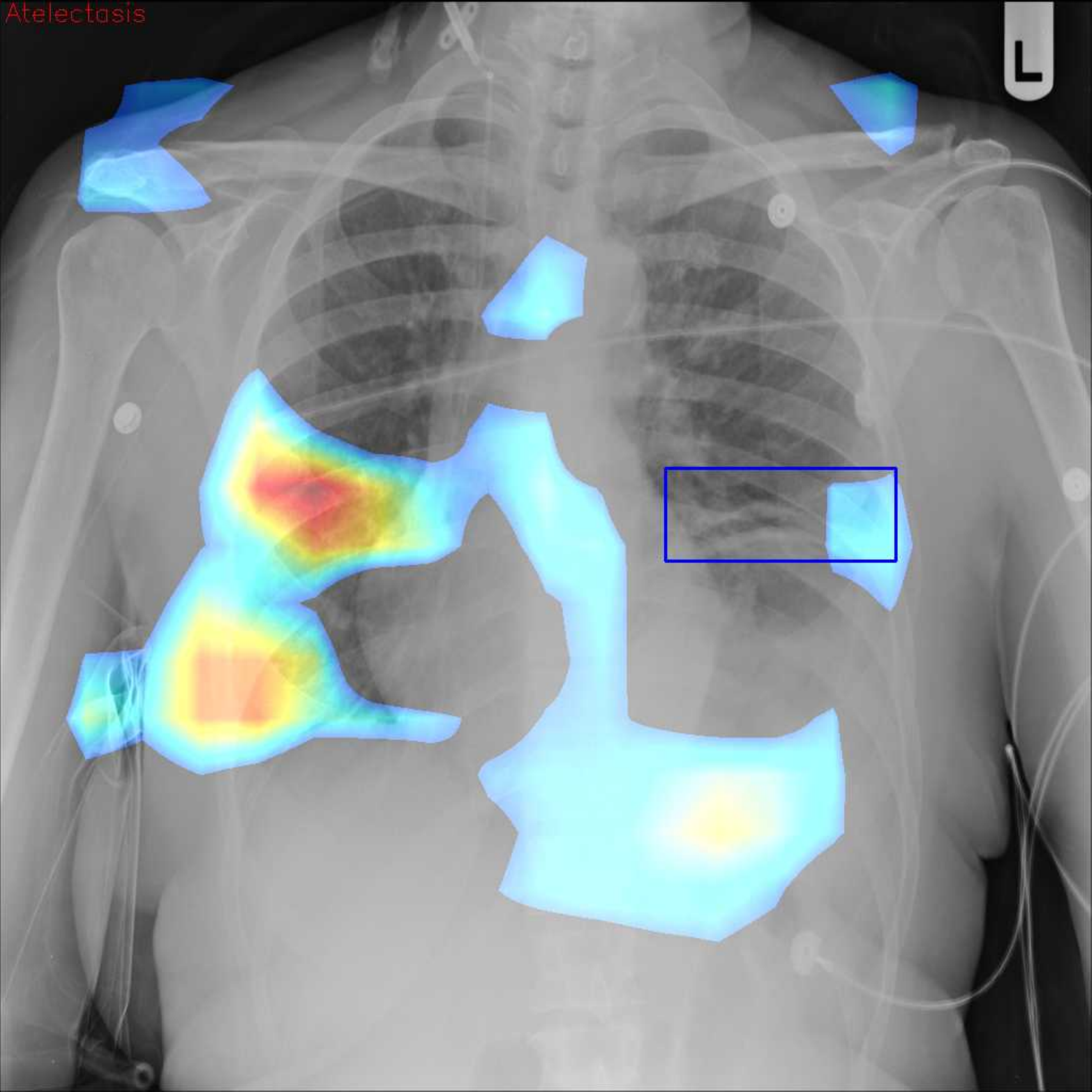} \\
           Cardiomegaly & \includegraphics[width=0.090\textwidth]{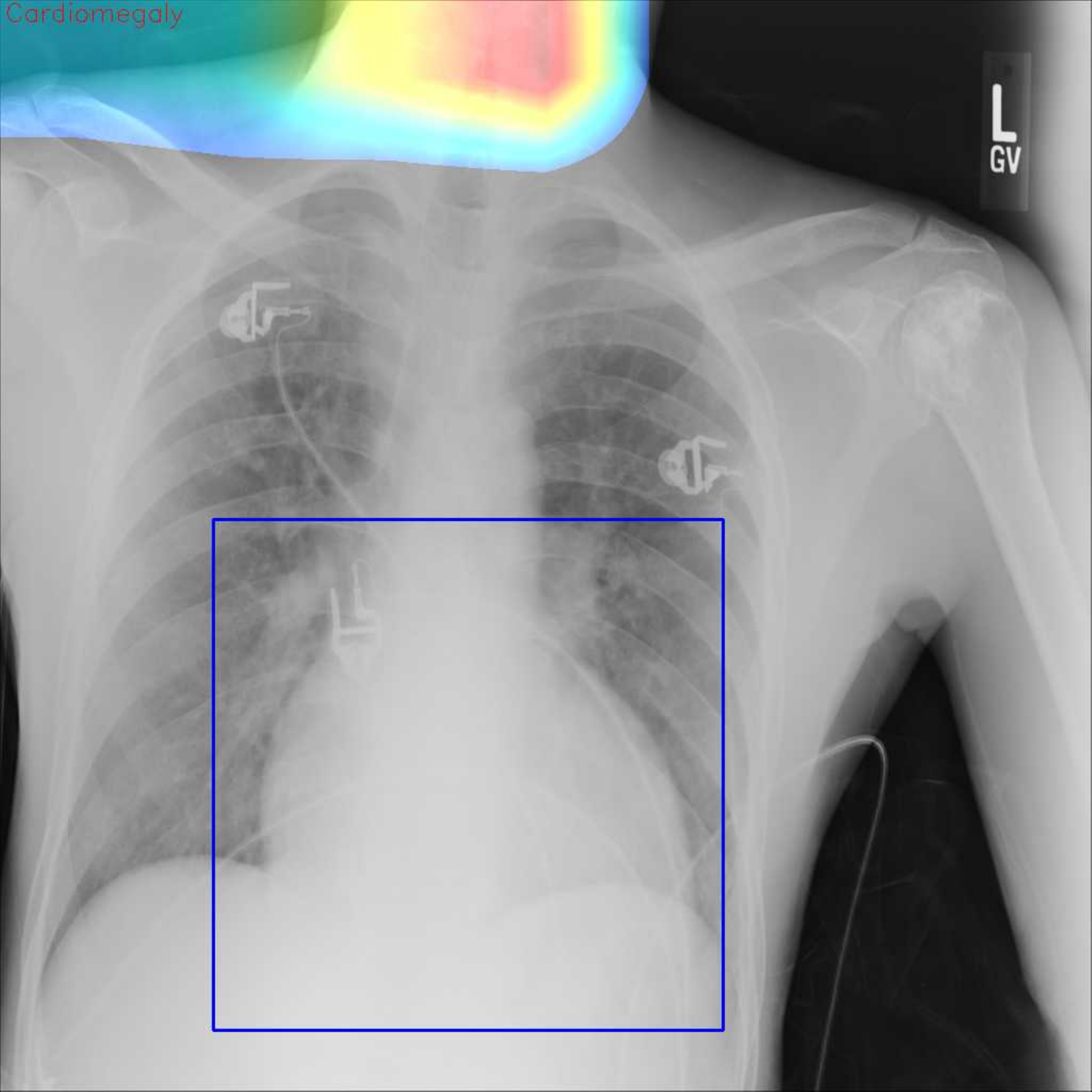} & \includegraphics[width=0.090\textwidth]{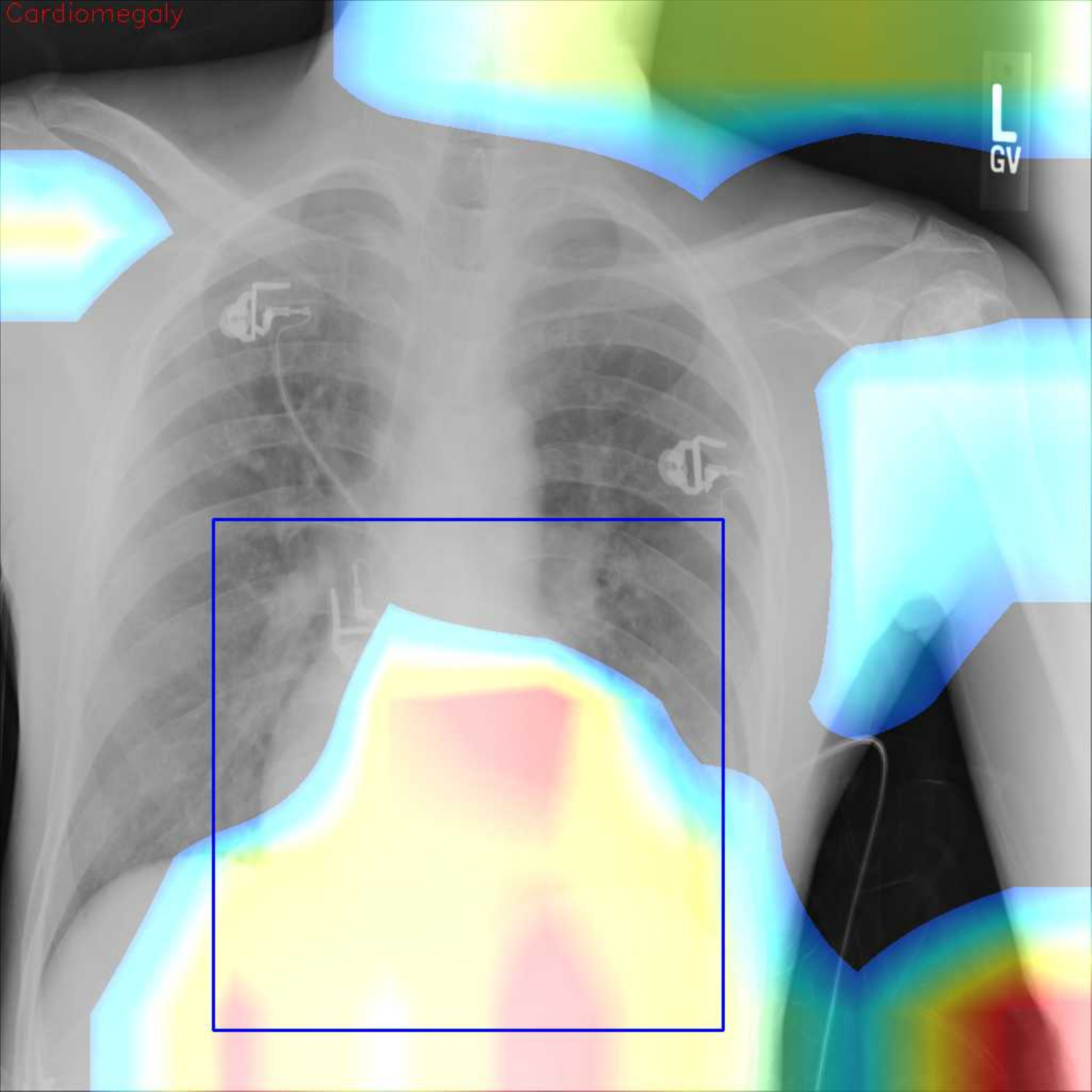} & \includegraphics[width=0.090\textwidth]{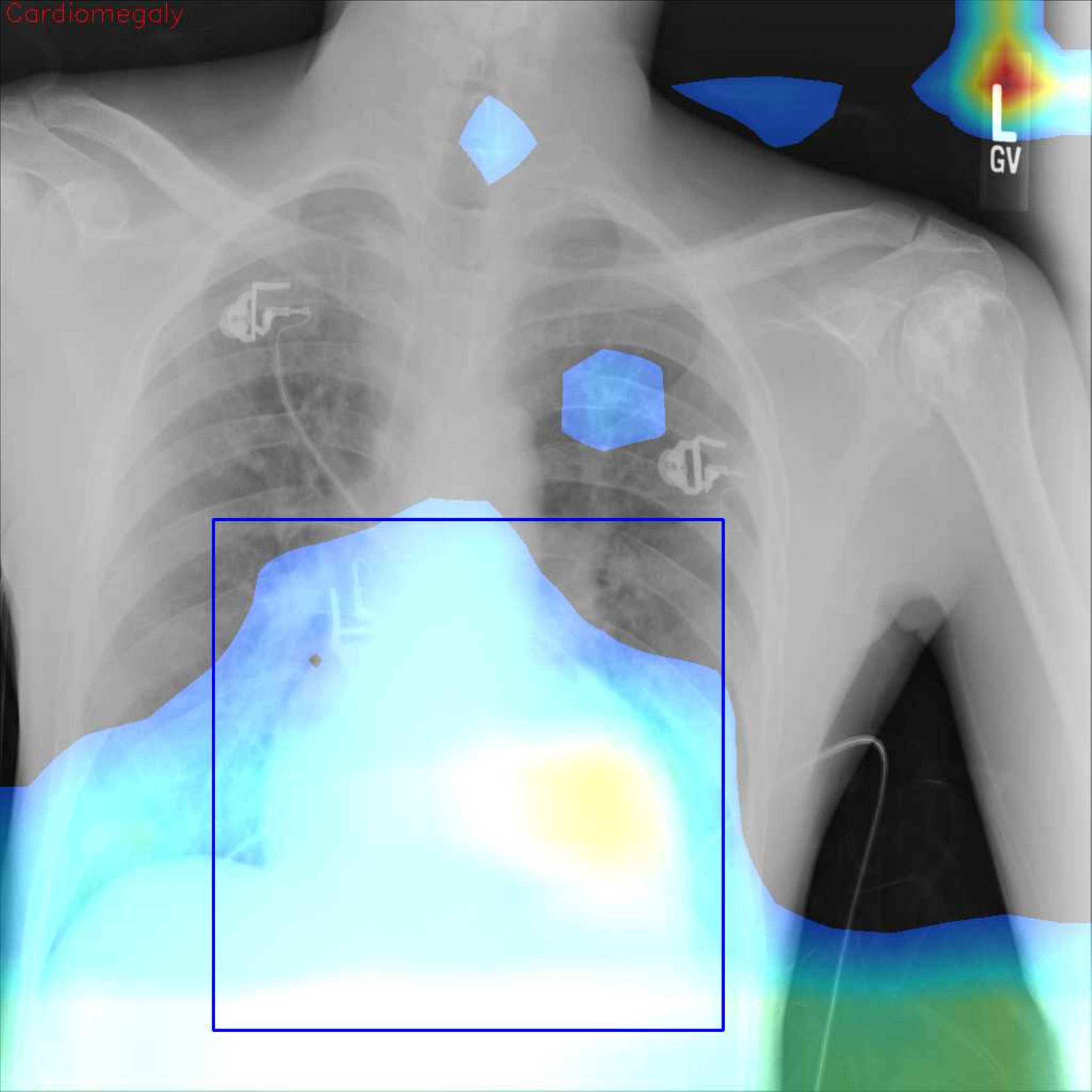} \\
           Effusion & \includegraphics[width=0.090\textwidth]{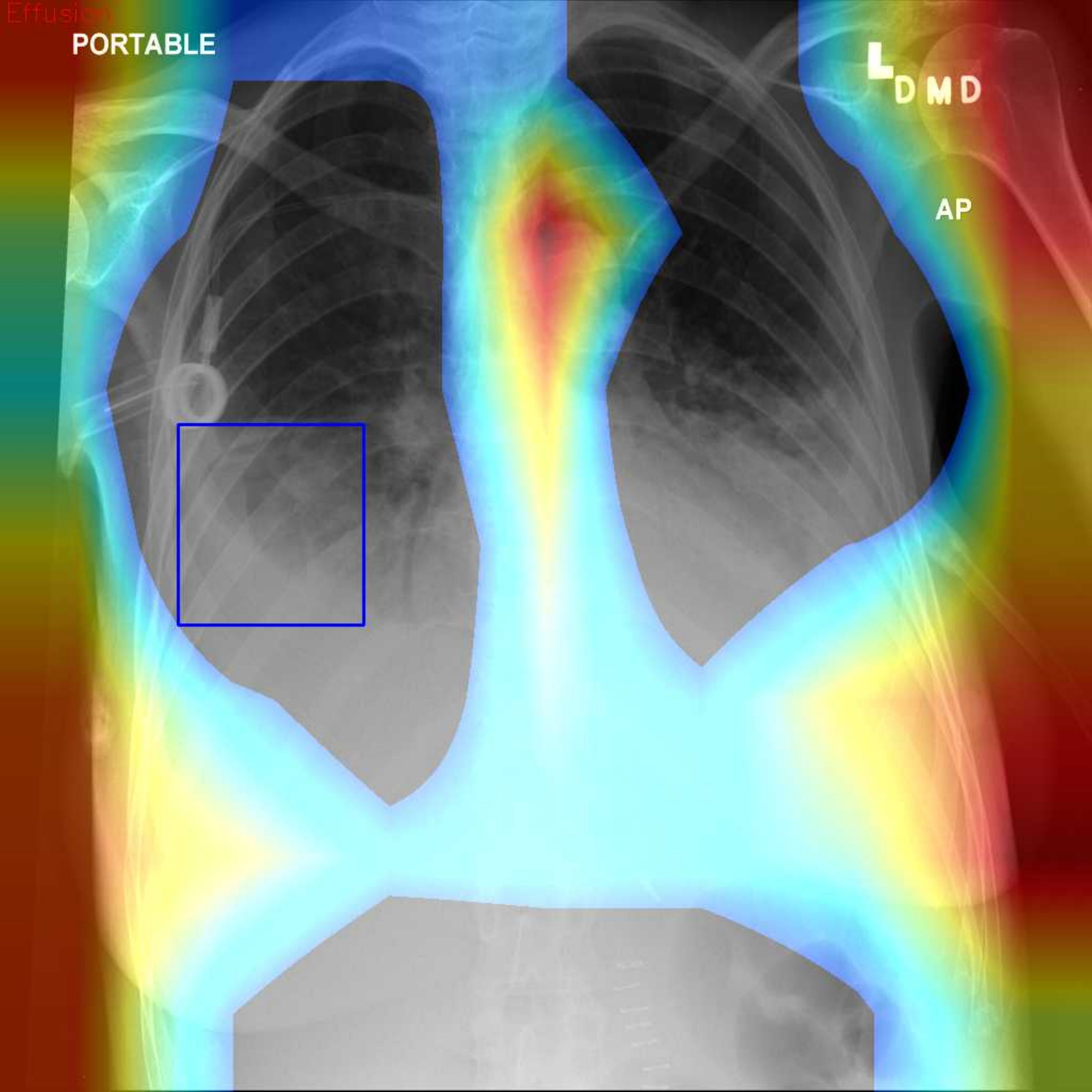} & \includegraphics[width=0.090\textwidth]{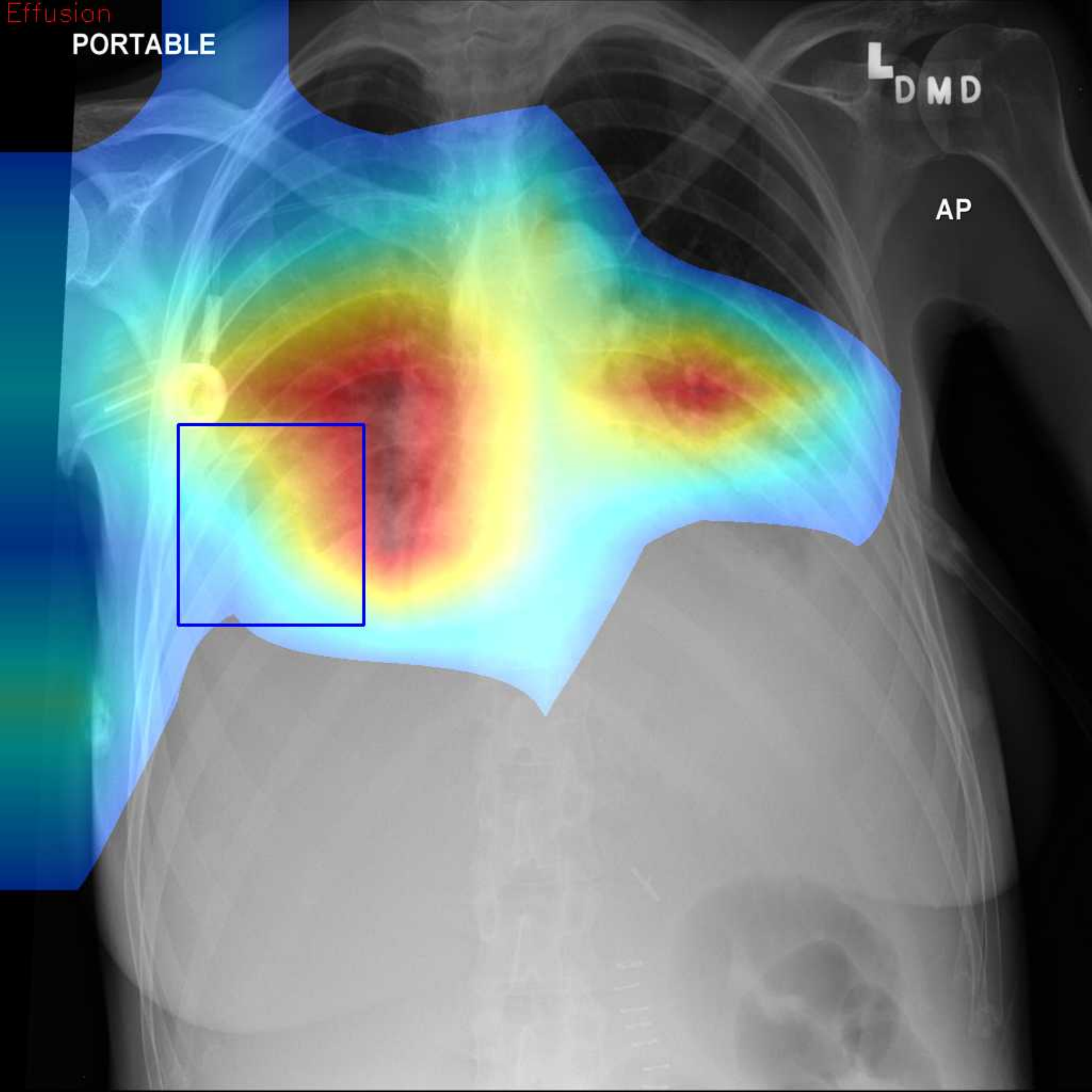} & \includegraphics[width=0.090\textwidth]{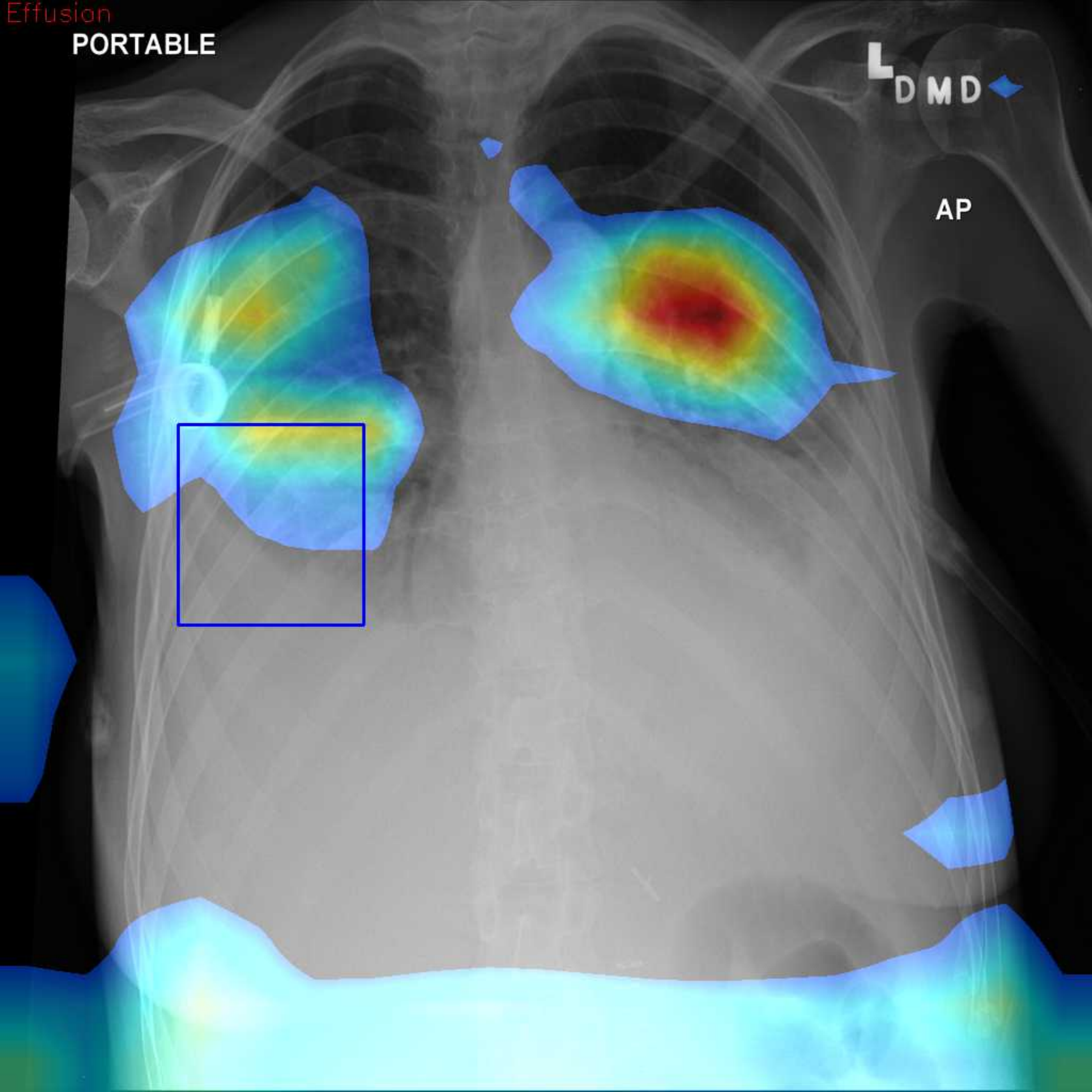} \\
          Infiltration & \includegraphics[width=0.090\textwidth]{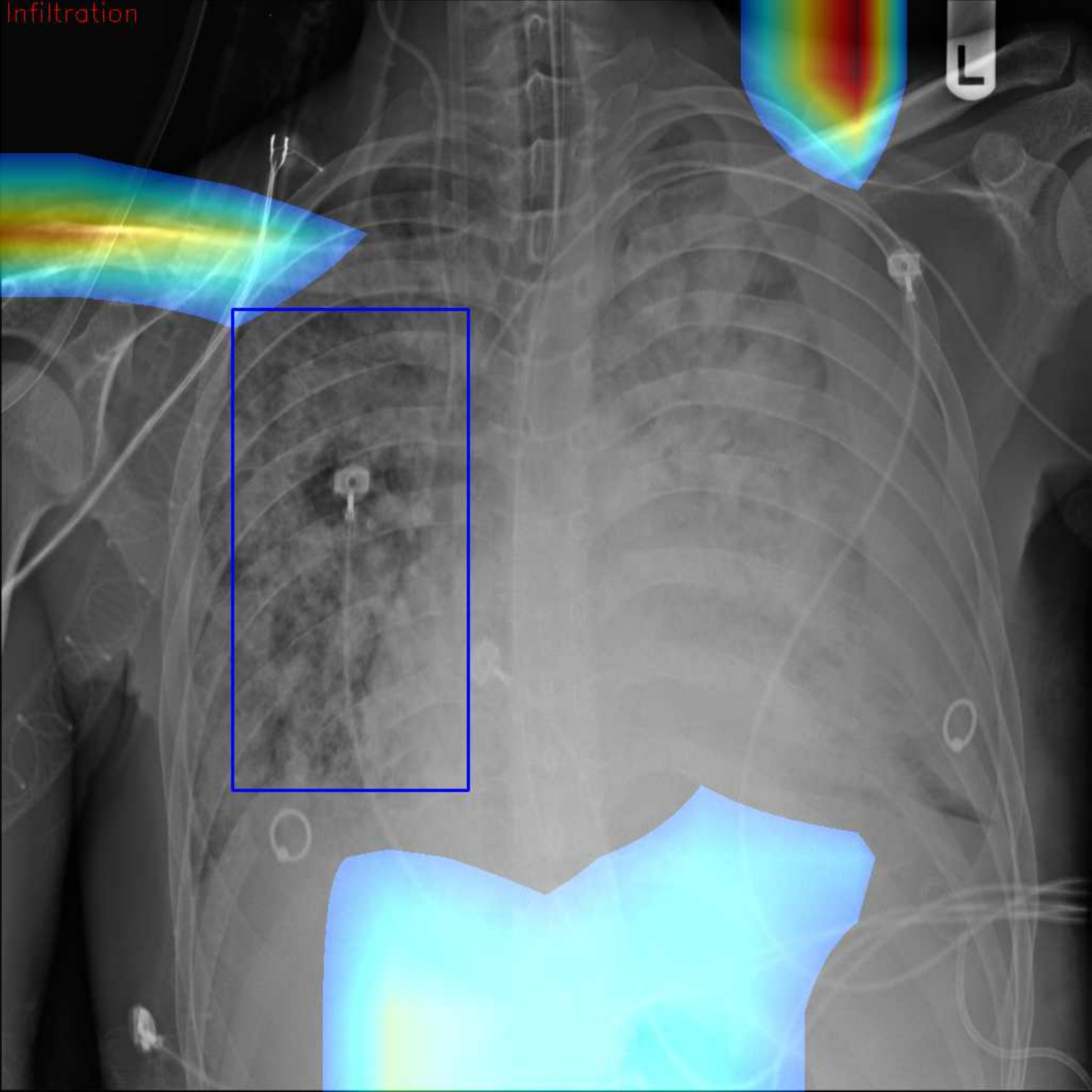} & \includegraphics[width=0.090\textwidth]{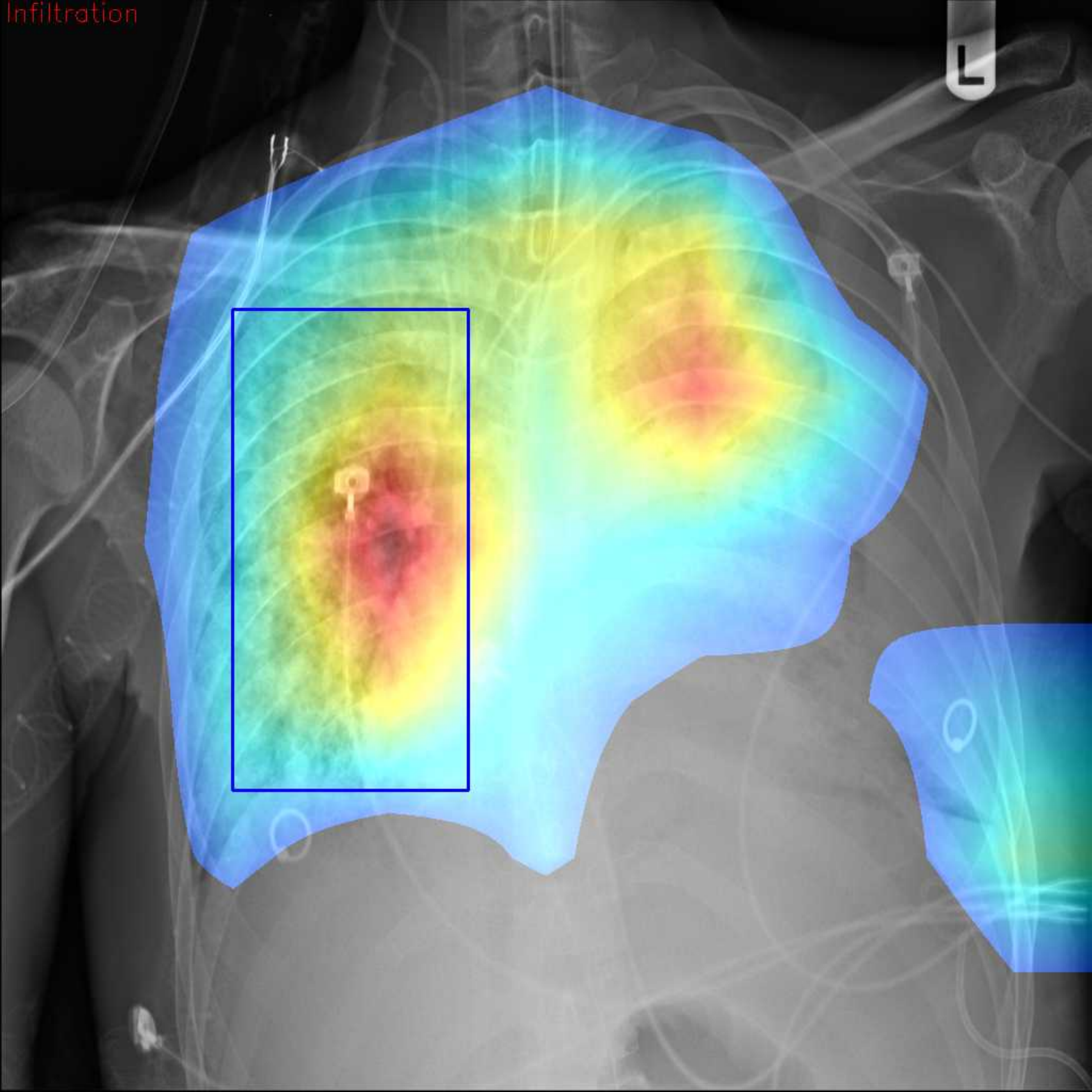} & \includegraphics[width=0.090\textwidth]{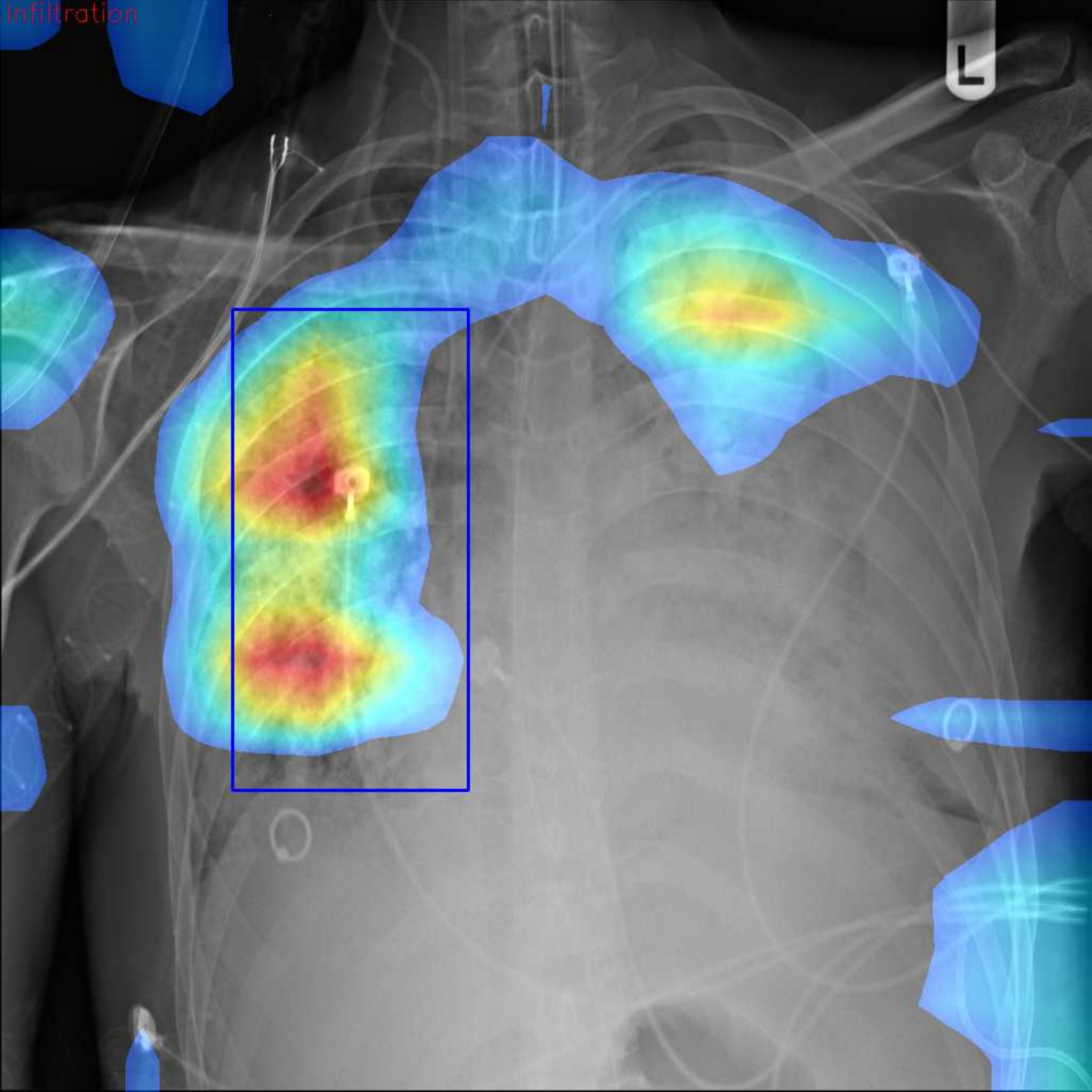} \\
           Mass & \includegraphics[width=0.090\textwidth]{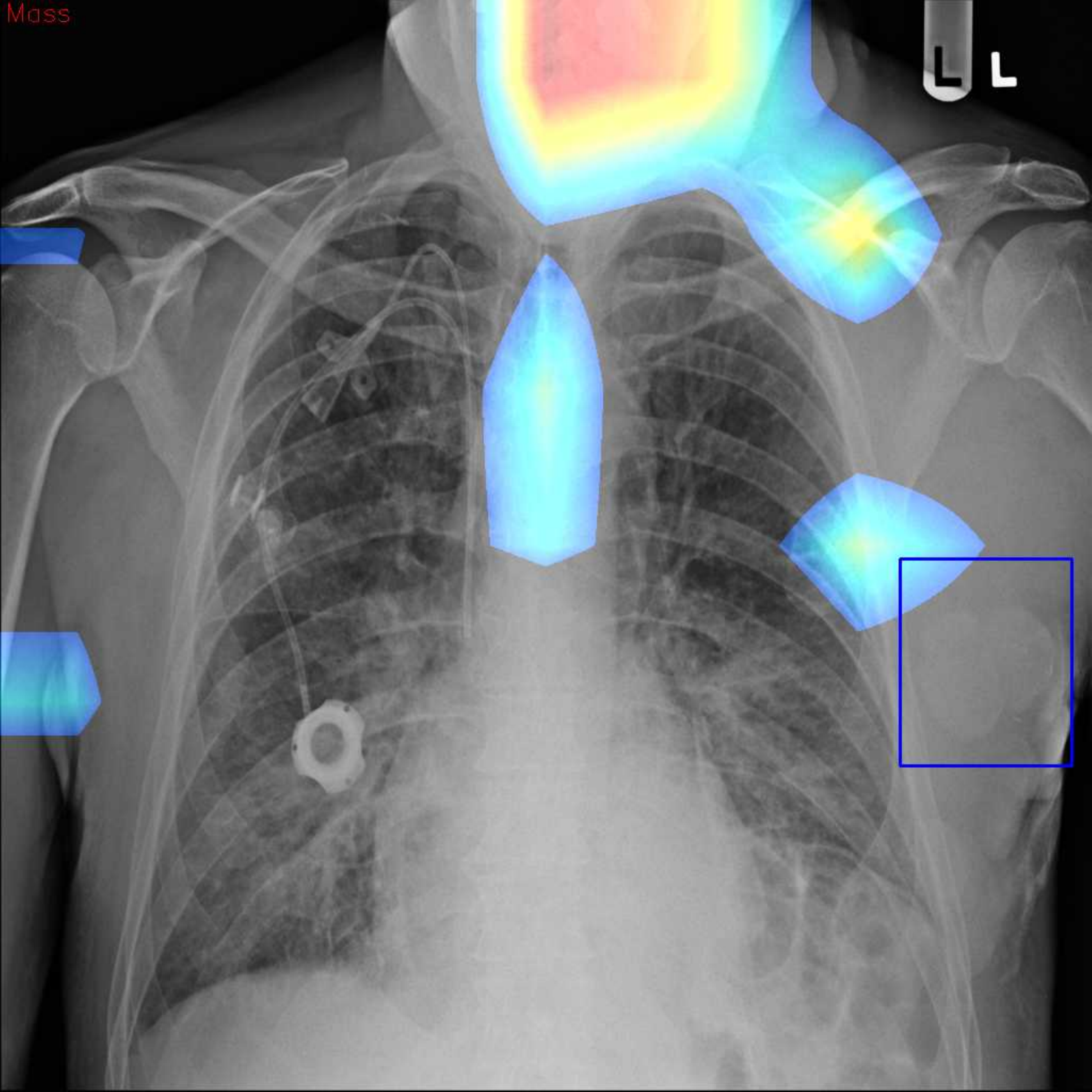} & \includegraphics[width=0.090\textwidth]{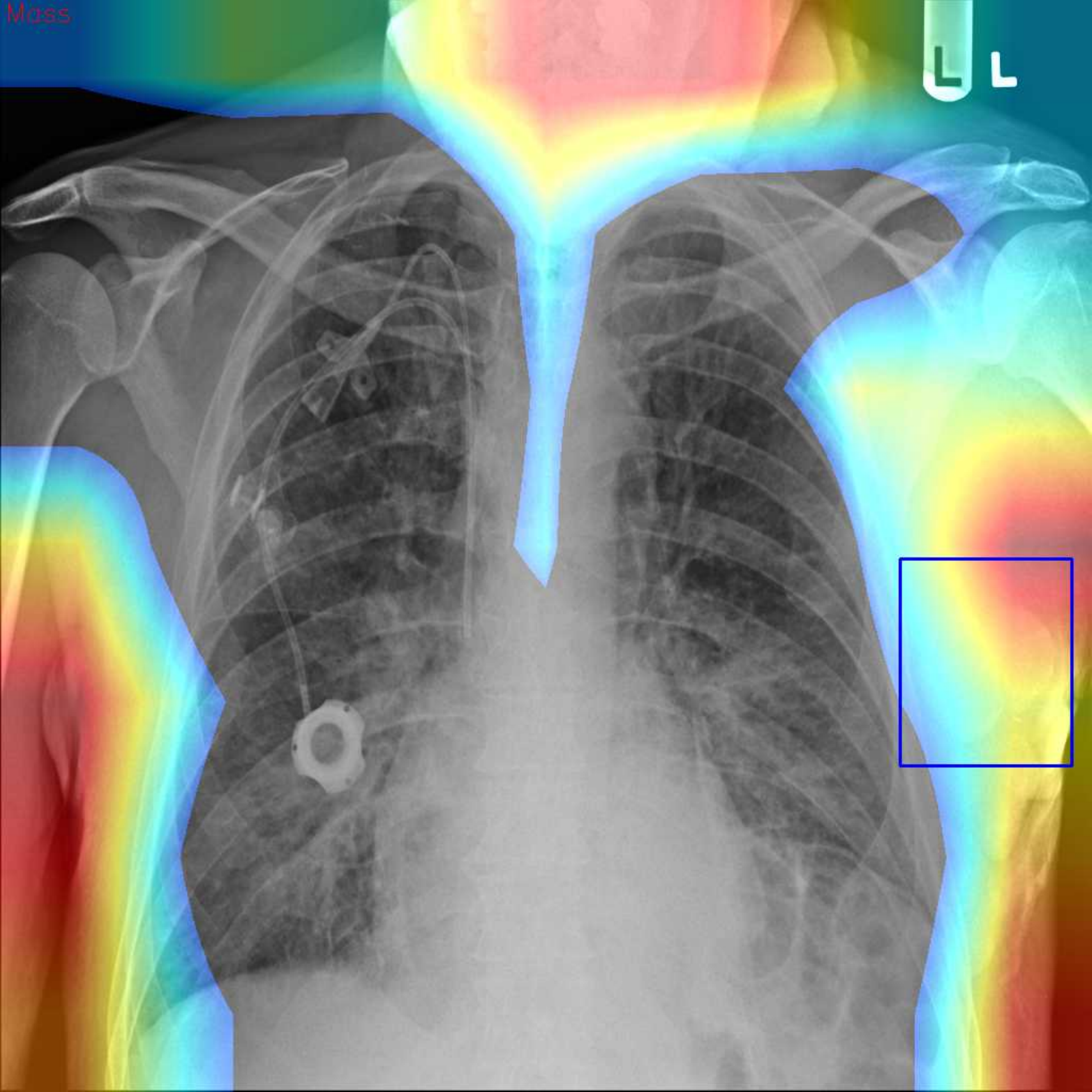} & \includegraphics[width=0.090\textwidth]{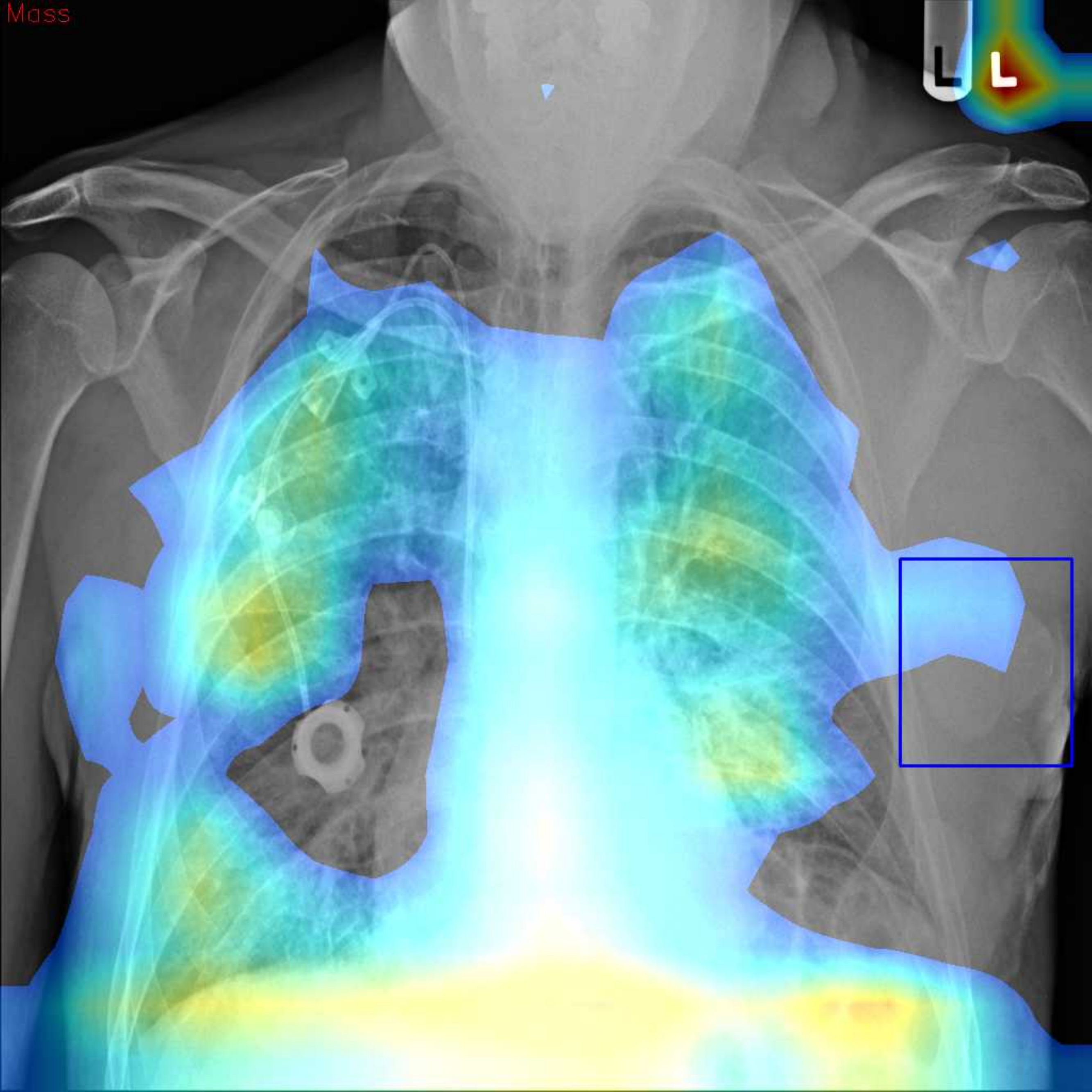} \\
            Nodule & \includegraphics[width=0.090\textwidth]{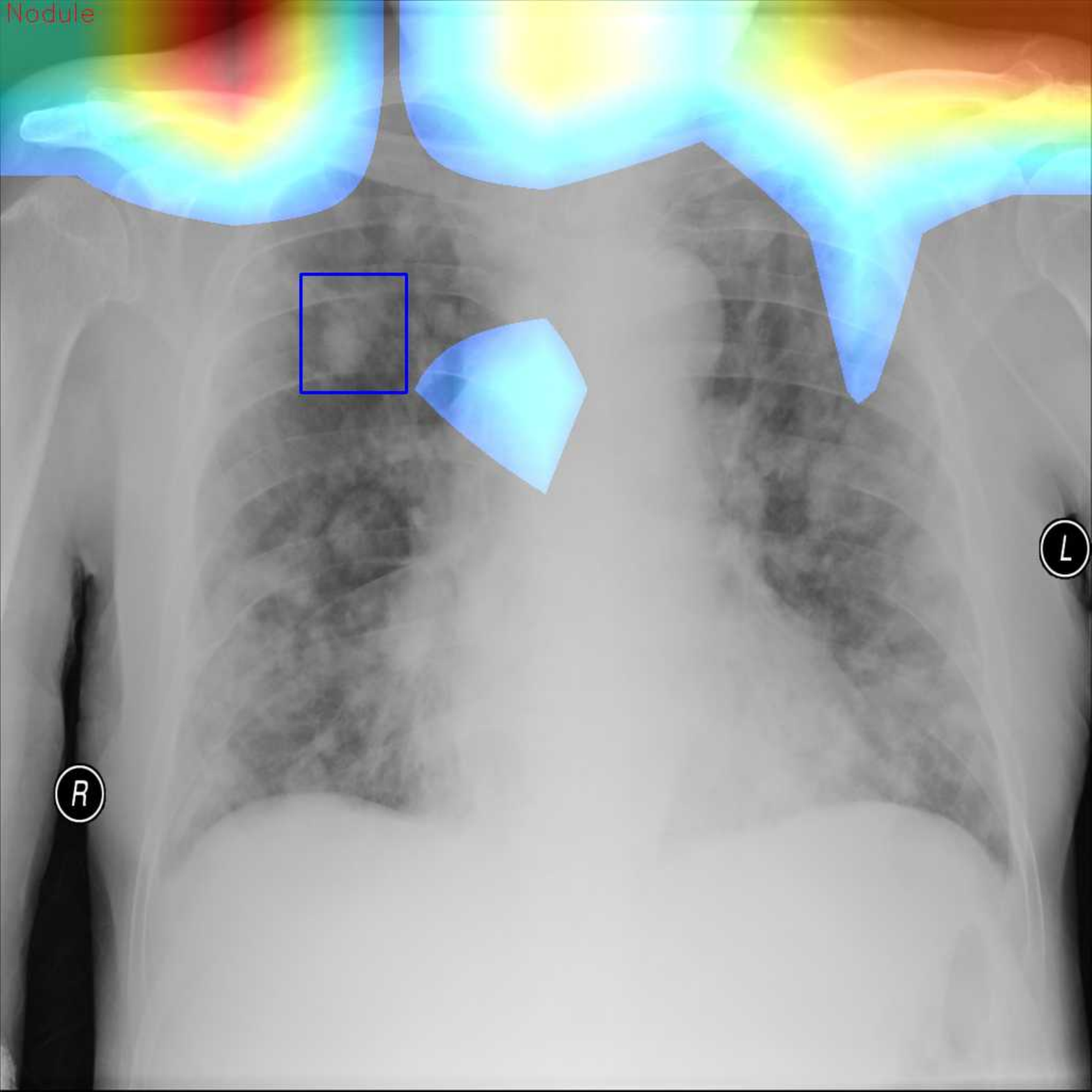} & \includegraphics[width=0.090\textwidth]{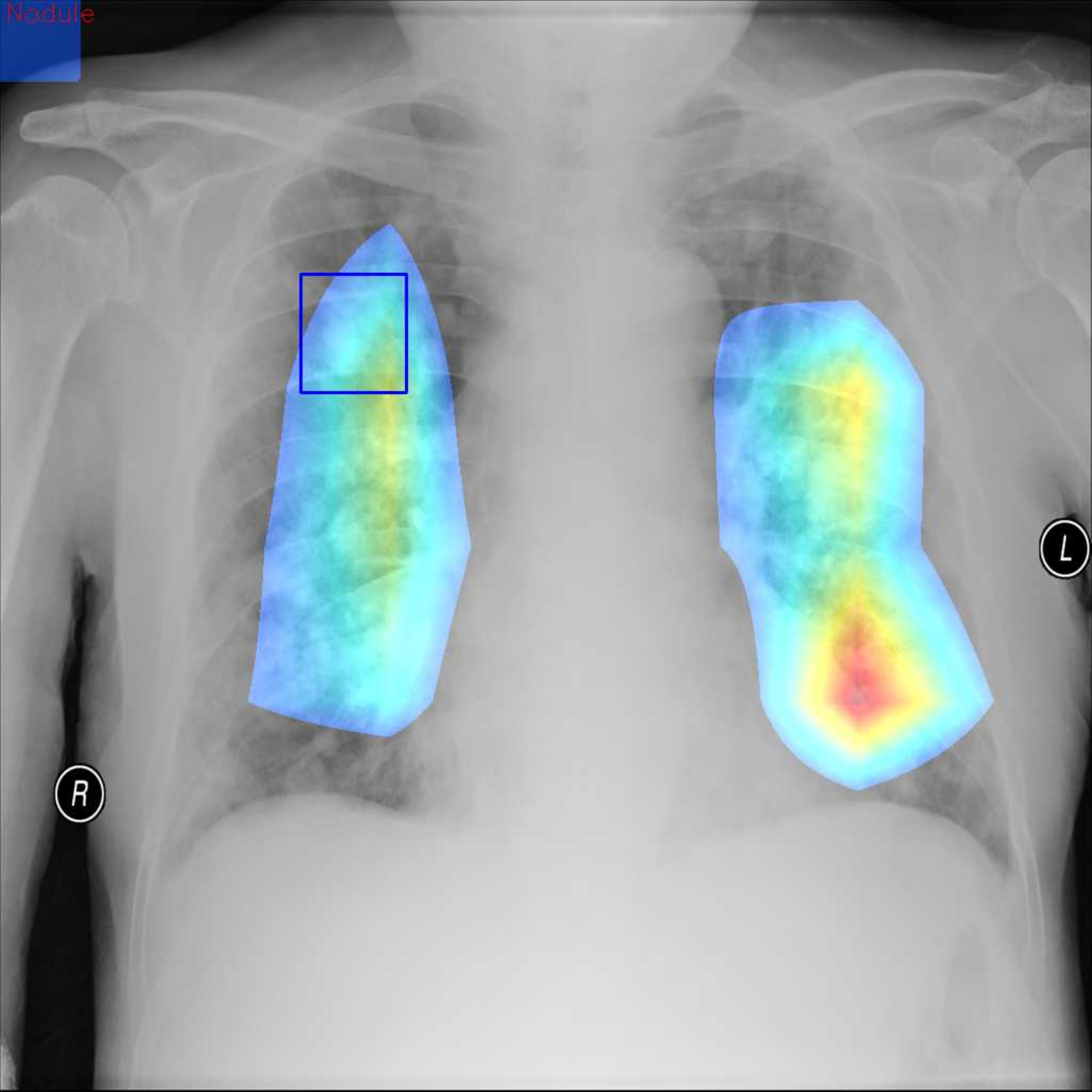} & \includegraphics[width=0.090\textwidth]{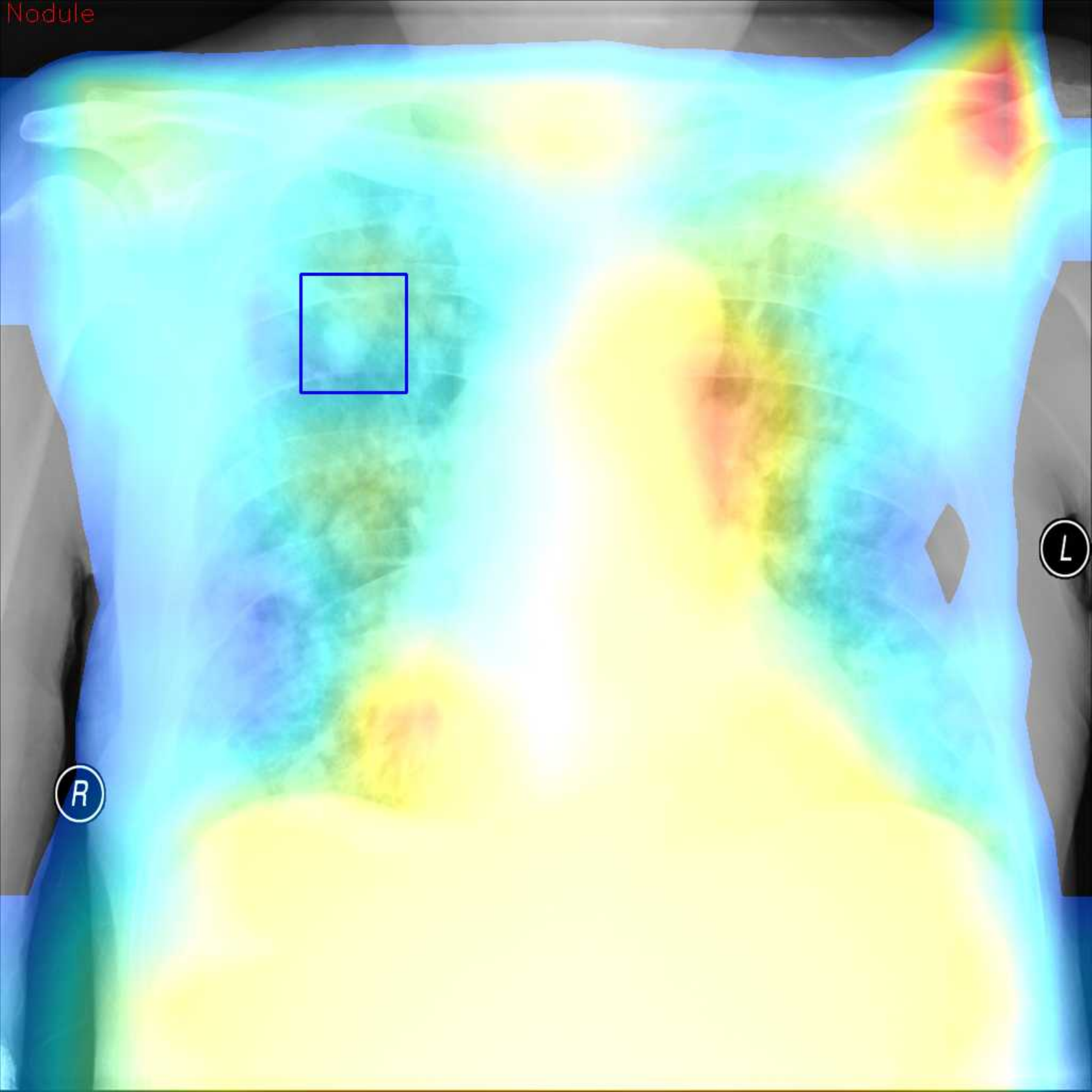} \\
            Pneumonia & \includegraphics[width=0.090\textwidth]{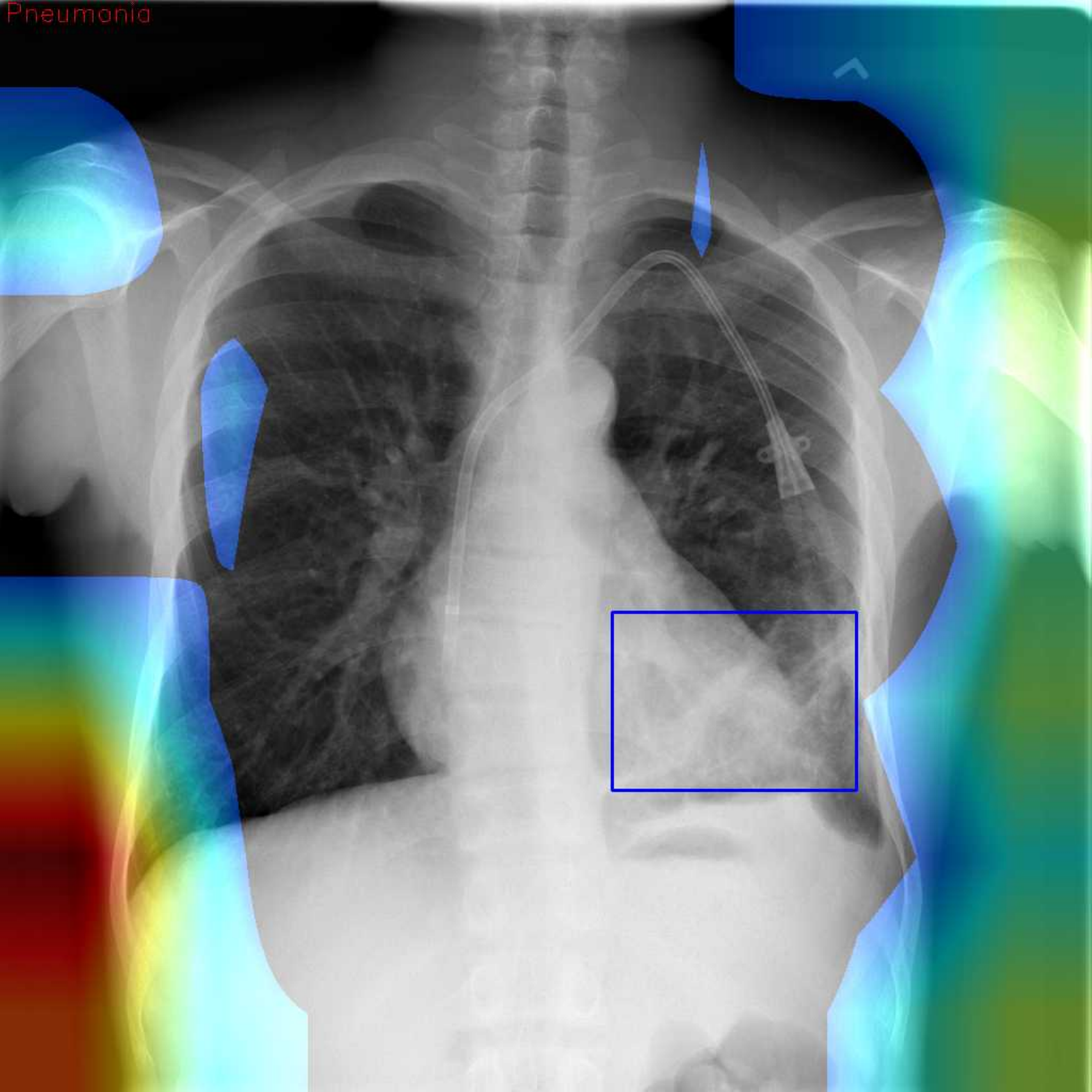} & \includegraphics[width=0.090\textwidth]{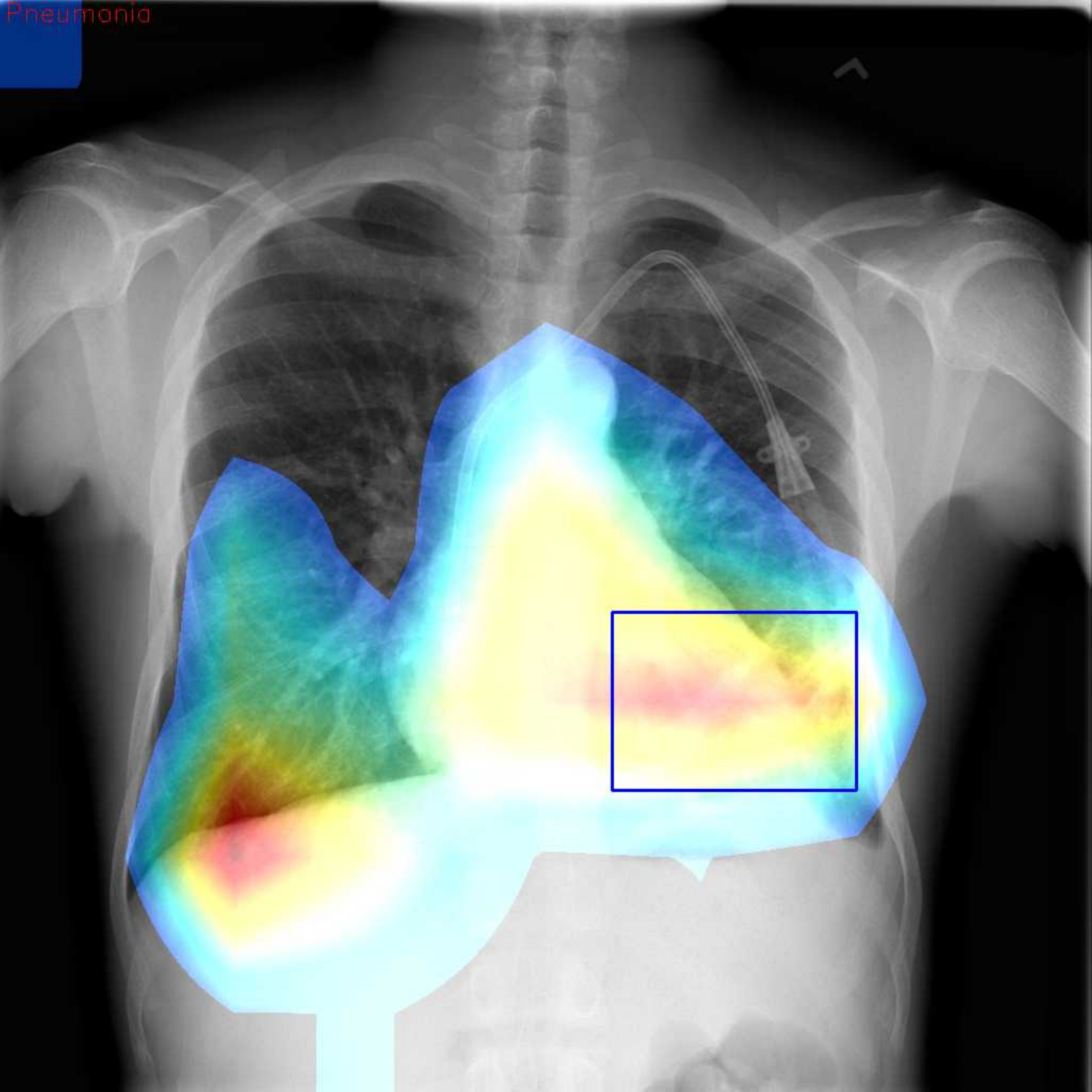} & \includegraphics[width=0.090\textwidth]{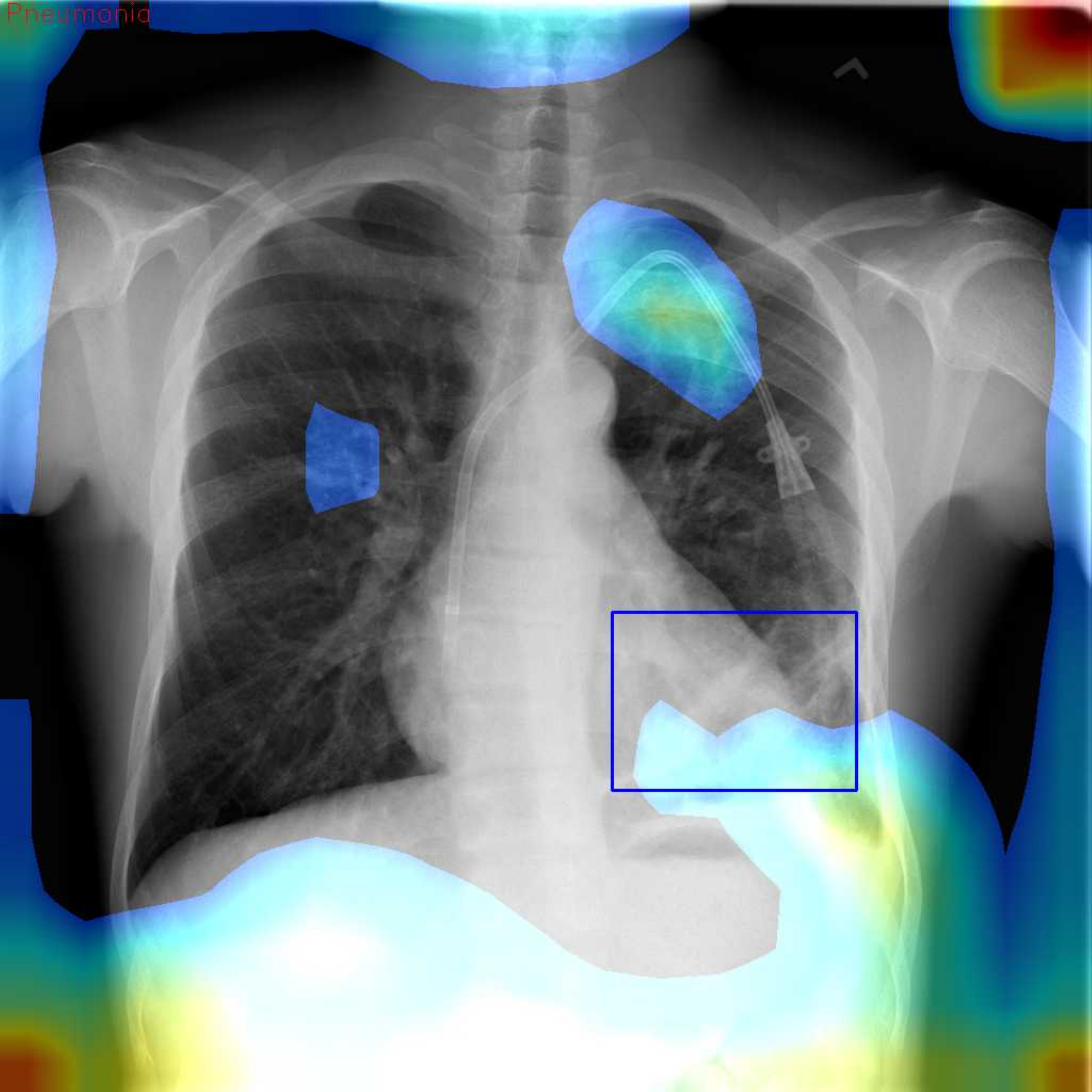} \\
            Pneumothorax & \includegraphics[width=0.090\textwidth]{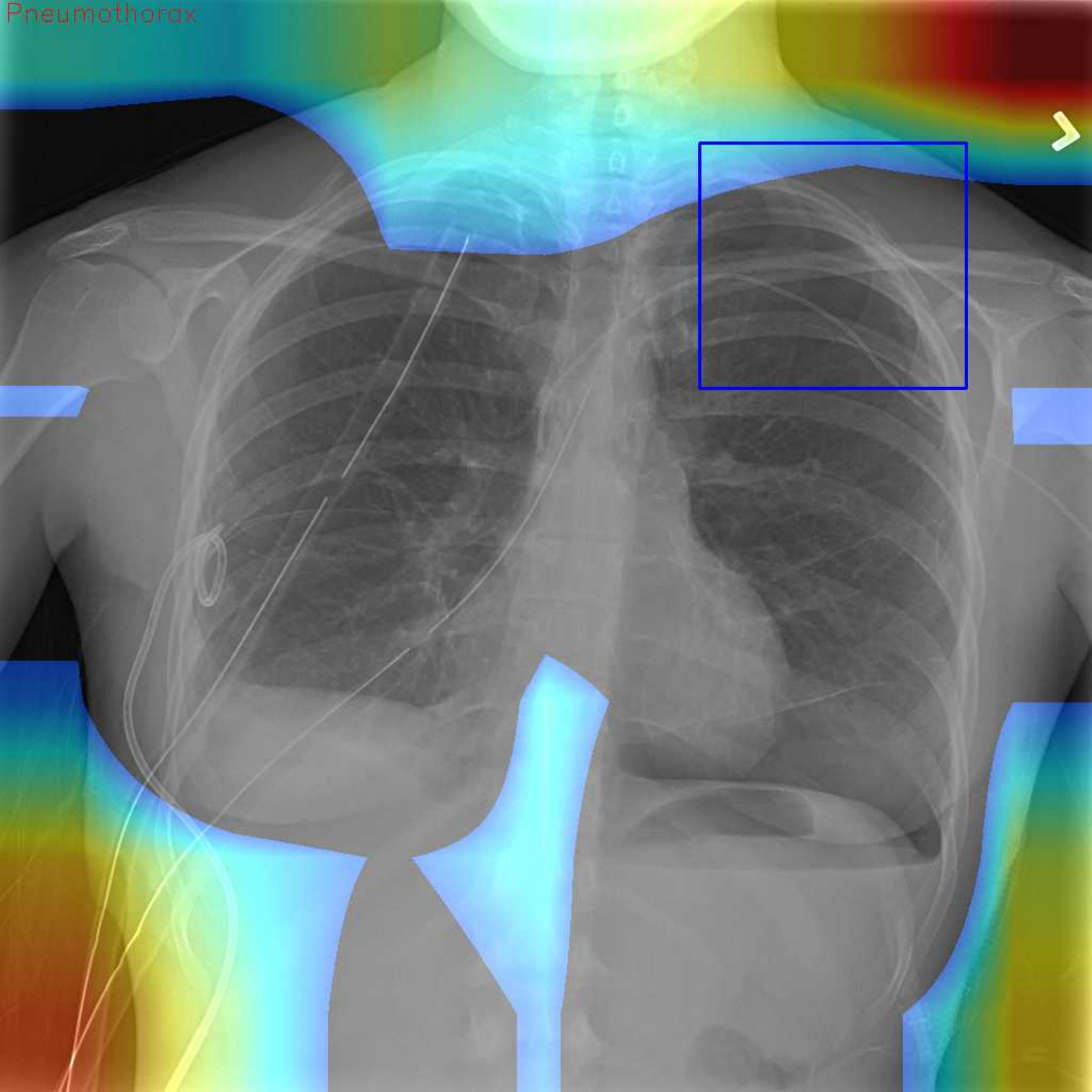} & \includegraphics[width=0.090\textwidth]{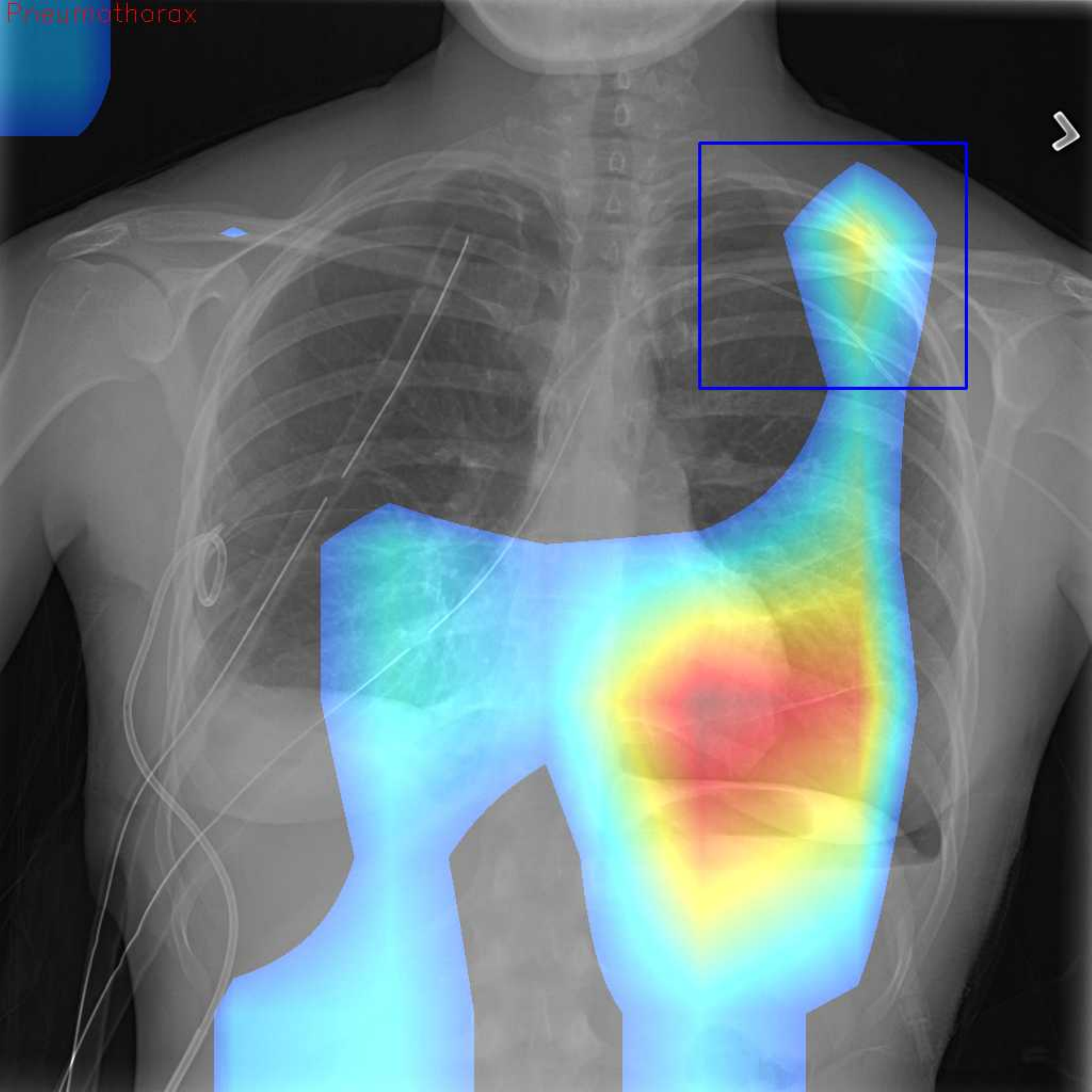} & \includegraphics[width=0.090\textwidth]{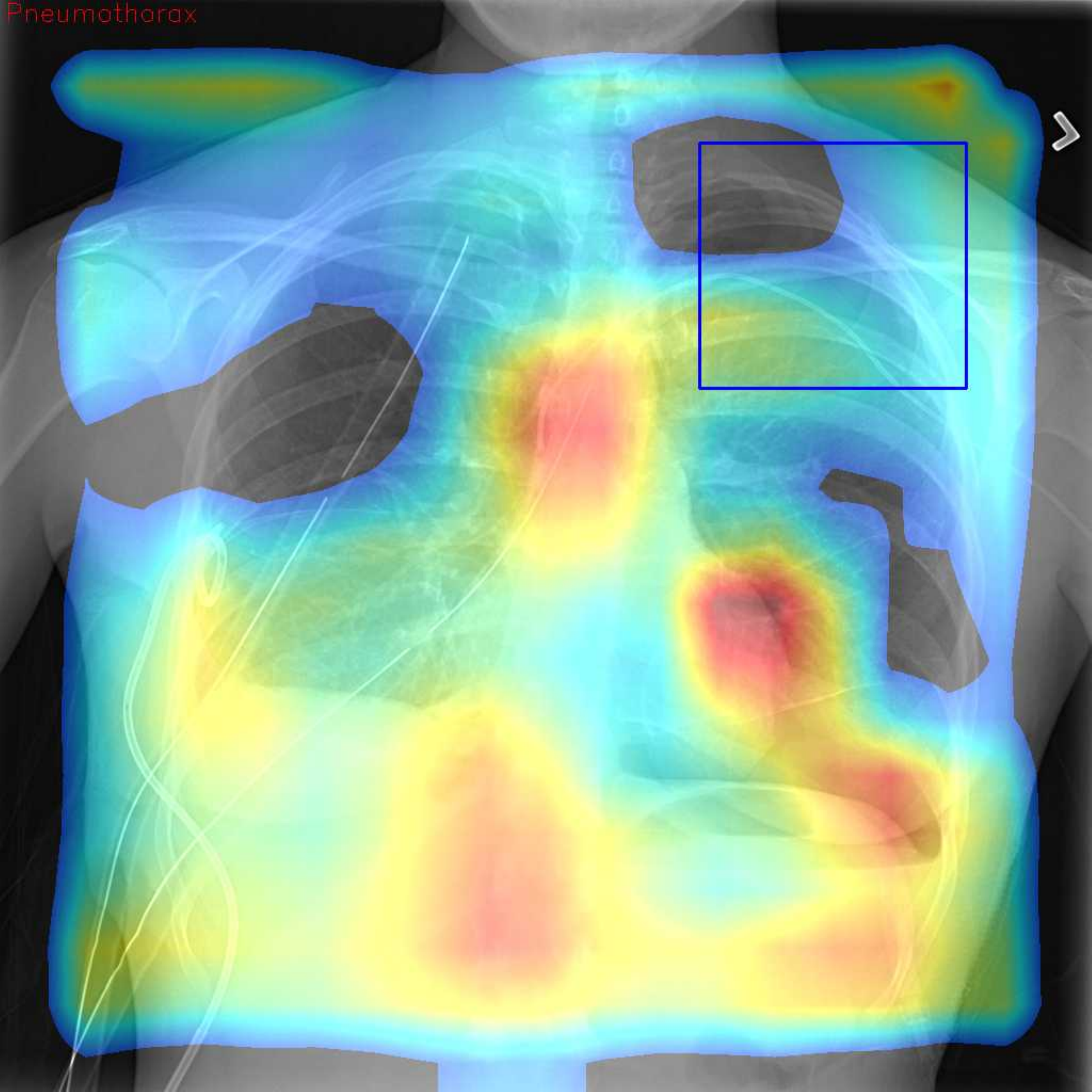} \\
            \bottomrule
        \end{tabular}
        }
\end{table*} 

\pdfbookmark[section]{Conclusion}{Conclusion}
\section*{Conclusion}
We proposed an aggregate of novel weighting function to formula the focal-loss function in complement with the two-stage training of EfficientNet, a state-of-the-art neural network architecture. We aim to improve the classification capability. Existing approaches of weighting function did not address the sample characteristics of both the positive-negative and easy-hard. The proposed weighting function attempts to improve the classification capability by address both the sample characteristics,which are ignored by the existing methods. The proposed approach provides a better decision boundary to the multiclass classification problem since the proposed approach addresses the imbalances of both positive-negative and hard-easy samples, also the use of recent network architecture scale-up the performances better. The proposed approach is able to improve the classification rates by $2.10\%$ than the latest research's outputs which is measured in the area under the receiver operating characteristic curve (AUROC). The proposed method also achieved state-of-the-art results under three distinct experiments setup, currently the results are the best improvements for the Chest X-Ray dataset being used. Since the proposed approach only addresses multiclass classification problem and multilabel classifications are not tackled, future research will target on multilabel problems. The proposed approach will be further validated
\FloatBarrier
\pdfbookmark[section]{Data Availability}{Data Availability} 
\section*{Data Availability} 
The dataset for this research is available publicly in the Chest X-ray NIHCC repository:\\ https://nihcc.app.box.com/v/ChestXray-NIHCC.
\pdfbookmark[section]{Code Availability}{Code Availability}
\section*{Code Availability} 
The code for to reproduce the research is available publicly at the repository:\\ https://github.com/bayu-ladom-ipok/weOpen

\bibliography{references}
\pdfbookmark[section]{Acknowledgements}{Acknowledgements}
\section*{Acknowledgements}
This work is supported and funded by Ministry of Religious Affair (MORA) Scholarship, Republic of Indonesia. We also very thankful for NIH Clinical Center for their publicly released dataset.
\section*{Author contributions statement}
B.A.N. performed the works for the manuscript.
\pdfbookmark[section]{Additional Information}{Additional Information}
\section*{Additional Information}
\subsection*{Competing interests}
B.A.N. is a MORA Scholarship recipient from Sunan Ampel Islamic State University Surabaya - Indonesia, and he is also a research student in Curtin University Bentley - Australia. The scholarship recipient's primary email addresses are: bayu@uinsby.ac.id and bayu.lecture@gmail.com. There are no potential conflict of interests for the manuscript.
\end{document}